\documentstyle[preprint,aps,epsf,psfig]{revtex}
\begin{document}
\draft

\title{Shell Model Monte Carlo Investigation of Rare Earth Nuclei}

\author{J. A. White and S. E. Koonin}
\address{W. K. Kellogg Radiation Laboratory, 106-38, California
Institute of Technology\\ Pasadena, California 91125 USA}
\author{D. J. Dean}
\address{Physics Division, Oak Ridge National Laboratory, P.O. Box 2008,
Oak Ridge, Tennessee 37831-6373 USA}
\address{and}
\address{Department of Physics and Astronomy, University of Tennessee,
Knoxville, Tennessee, 37996} 
\date{\today}
\maketitle
\tightenlines
\begin{abstract}
We utilize the Shell Model Monte Carlo (SMMC) method to study the
structure of rare earth nuclei.  This work  
demonstrates the first systematic ``full oscillator shell plus intruder''
calculations in such heavy nuclei.
Exact solutions of a pairing plus quadrupole hamiltonian
are compared with mean field and SPA approximations in several Dysprosium
isotopes from $A=152-162$, including the odd mass $A=153$.  Basic
properties of these nuclei at various temperatures and spin are
explored.  These include energy, deformation, moments of inertia,
pairing channel strengths, 
band crossing, and evolution of shell model occupation numbers.  Exact
level densities are also calculated and, in the case of $^{162}$Dy,
compared with experimental data.
\end{abstract}
\narrowtext
\section{Introduction}
\label{intro}
Our goal is to develop an improved
microscopic understanding of the structure of rare earth nuclei; i.e.,
an understanding based on the behavior of individual nucleons in the
nucleus.  Toward that end we solve the shell model systematically in a full
oscillator shell basis with intruders for the first 
time in rare earth nuclei using the Monte Carlo (SMMC) technique;
calculations using other methods have been restricted to a severely
truncated model space.  SMMC allows us to trace structural
rearrangements within nuclei induced by changes in temperature and
spin, so that we may obtain a clearer microscopic picture of general
structural features in this region of the periodic table.

We assume an effective two-body nucleon-nucleon
interaction and perform a Hubbard-Stratonovich transformation to obtain a path
integral representation for the partition function, which is then
evaluated by Monte Carlo methods (see Section~\ref{smmc}) to
produce an exact shell model solution within statistical errors; this
substantially enhances the predictive power  
of the nuclear shell model for some observables.  Indeed, direct
diagonalizations of the shell model Hamiltonian in a full basis have
been limited to A$\sim50$, while we present calculations for A$\sim150$.

We examine how the phenomenologically motivated ``pairing plus quadrupole''
interaction compares in exact shell model solutions vs. the mean field
treatment.  We also examine how the shell model solutions compare with
experimental data;  static path approximation (SPA)
calculations are also shown.  There have been efforts 
recently by others to use SPA calculations, since it is simpler and
faster (see \cite{Rossignoli,Rossignoli2} as examples).
However, the SPA results are not consistently good.  In particular, it
is useful to know not only if 
phenomenological pairing plus quadrupole type interactions can be used
in exact solutions for large model spaces, but also if the parameters
require significant renormalization because this affects the accuracy
of the SPA. 

We study a range of Dysprosium isotopes (Z=66, $86\leq N\leq96$),
which exhibit a  
rich spectrum of the behaviors such as shape
transitions, level crossings, and pair transfer that have been
observed in the rare earths.  These results should
therefore
apply quite generally in the rare earth region, although the immediate work
focuses on Dysprosium.  We have selected this element since the
half-filled proton shell makes the model spaces particularly large.

A previous paper discussed SMMC for the test case $^{170}$Dy, which
does not exist as a stable nucleus~\cite{Dy170}.  The work presented
here is much more systematic and thorough.  Algorithm improvements
subsequent to ~\cite{Dy170} have increased the computational execution
by a factor of 10 or more and have allowed us to calculate the rare
earths at lower temperatures, nuclear shapes are calculated using the
correct calculated quadrupole variance (not just a constant), and
pairing operators not used in ~\cite{Dy170} are calculated.

\section{Theoretical background}
\label{smmc}
Shell model diagonalization is still limited to $A\sim 50$ in the 0f1p
shell~\cite{Caurier}.  In 
contrast, SMMC determines thermal observables, but explicit wave functions are
never constructed; this is the key to how the
predictive power of the shell model is extended so tremendously.  The method 
is far less demanding on machine storage and there is no need to perform
manipulations with the exponentially increasing numbers of variables that are
encountered in direct diagonalization.
SMMC storage scales like
$N_s^2 N_t$, where $N_s$ is the number of single particle shell model
states and $N_t$ is the number of time slices (see below).

No known discrepancies exist between SMMC and direct diagonalization
in cases where the comparison has been possible.  This includes odd
mass nuclei computed for appropriate temperatures. 
Realistic fp and sd-shell solutions using modified KB3 and
Brown-Wildenthal interactions, respectively, agree with 
experiments~\cite{kdl}.  These
results give us a high degree of confidence in the SMMC technique.

As with any shell model, an effective nucleon-nucleon interaction must be
specified.  We use the well-known pairing plus quadrupole interaction as
formulated by Kumar and Baranger~\cite{Kumar-Baranger}.  The
Hamiltonian is
\begin{equation}
\hat H=\hat H_{sp}-G_p\hat P_p^\dagger\hat P_p-G_n\hat P_n^\dagger\hat P_n
-\frac{\chi}{2}\hat Q\cdot\hat Q \label{eq:ham}
\end{equation}
with $\hat Q=\hat Q_p+\hat Q_n$.  The pairing and quadrupole operators
are defined as 
\begin{eqnarray}
\hat P^\dagger_{J=0}&=& \sum_{jm} (-)^{j-m+l}\hat a^\dagger_{jm} \hat a^\dagger_{j-m} \\ 
\hat Q\cdot\hat Q &=&\sum_{i,j,k,l} \langle i|Q_\mu|k\rangle \langle l|Q_\mu | j\rangle \hat a^\dagger_i \hat a^\dagger_j \hat a_l \hat a_k, 
\end{eqnarray}
where $Q_\mu=r^2 Y_{2\mu}(\theta,\phi)$ as usual.  The single particle
energies are also taken from Kumar and Baranger~\cite{Kumar-Baranger}.

Effective charges are incorporated to account for
core polarization to 
fit measured electric quadrupole transition strengths.  The electric
quadrupole operator, with effective charges $e_p$ and $e_n$, is
\begin{equation}
\hat Q=e_p \hat Q_p+e_n \hat Q_n
\end{equation}

\subsection{Method and sign problem}
\label{sign}
Detailed procedures for the SMMC are explained
fully in~\cite{kdl} and references therein.  We provide no further
explanation, except as regards the ``sign problem.''  

We define $\Phi\equiv$Tr$\hat U_\sigma/|$Tr$\hat U_\sigma|$ as the sign for a 
given Monte Carlo sample, where $\hat U_\sigma$ is defined as 
\begin{eqnarray}
\hat U_\sigma &=& \hat U_{N_t}\hat U_{N_t-1}...\hat U_1 \\
\hat U_n &=& e^{-\Delta\beta\hat h_\sigma}  \\
\end{eqnarray}
$\hat h_\sigma$ is the one body hamiltonian for the auxiliary field
configuration $\sigma$ and $\beta$ is the inverse temperature. 
In SMMC, the partition function path integral is divided
into $N_t$ time steps of size $\Delta\beta$ so that
$\beta=N_t\Delta\beta$.  Hence, the 
complete evolution operator is 
expressed as a product of operators in each time step.  In these
studies, we use the canonical (number projection) formalism to
evaluate the trace. 

If $\Phi$ is not equal to one, 
numerical instabilities can arise.  This has become widely known as
the Monte Carlo ``sign problem.''  The simple phenomenological pairing
plus quadrupole interaction (without added pn pairing) does not have
an inherent sign problem.  However, 
sign problems can arise even with this simple interaction if time
reversal symmetry is broken, as when odd masses are studied or the
system is cranked by adding a term $-\omega \hat J_z$ to $\hat H$.  In
these studies, the sign violation turned out to be  
minor for odd mass ground states and canonical ensemble cranking was
limited to $\omega\leq0.3$ MeV.  Experimentally, these nuclei are
observed to $\omega\approx0.6$ MeV~\cite{pckhoo}. 

\subsection{Shapes and moments of inertia}
The quadrupole expectation values $\langle Q_\mu \rangle$ vanish under
rotational symmetry.  However, for a given Monte Carlo sample $Q_\mu$
will have some finite, nozero value.  We calculate $Q_{ij}=3x_i
x_j-\delta_{ij}r^2$ for each sample and relate its eigenvalues to the
quadrupole $\beta$ and $\gamma$ deformation parameters as done previously with
SMMC \cite{ormandq,Dy170}.  The intrinsic frame for each sample with a
field configuration $\sigma$ has nonzero components $Q'_0$ and
$Q'_2=Q'_{-2}$ as 
\begin{eqnarray}
\langle Q_{0}'\rangle_\sigma &=& \frac{3}{2\pi} \sqrt{\frac{4\pi}{5}}
\langle r^2 \rangle_\sigma \beta_\sigma \cos\gamma_\sigma \nonumber\\
\langle Q_{2}'\rangle_\sigma &=& \frac{3}{2\pi} \sqrt{\frac{4\pi}{5}}
\langle r^2 \rangle_\sigma \frac{\beta_\sigma}{\sqrt{2}}
\sin\gamma_\sigma
\end{eqnarray}
In terms of eigenvalues $Q'_{11}, Q'_{22}$, and $Q'_{33}$ of $Q_{ij}$ give
\begin{eqnarray}
\langle Q_{11}'\rangle_\sigma &=& \sqrt{\frac{2\pi}{5}}
\left(\sqrt{3}(\langle Q_{2}'\rangle_\sigma +\langle  
Q_{-2}'\rangle_\sigma)
-\sqrt{2}\langle Q_{0}'\rangle_\sigma \right)\nonumber \\
\langle Q_{22}'\rangle_\sigma &=& \sqrt{\frac{2\pi}{5}}
\left(-\sqrt{3}(\langle Q_{2}'\rangle_\sigma +\langle  
Q_{-2}'\rangle_\sigma)
-\sqrt{2}\langle Q_{0}'\rangle_\sigma\right) \nonumber \\
\langle Q_{33}'\rangle_\sigma &=& 2\sqrt{\frac{4\pi}{5}}
\langle Q_{0}'\rangle_\sigma \nonumber
\end{eqnarray}
We can also calculate the free energy $F(\beta,\gamma)$ to construct
shape contour plots.  This is done using
\begin{equation}
F(\beta,\gamma)=-T\ln{\frac{P(\beta,\gamma)}{\beta^3\sin{3\gamma}}}
\label{shapeeq}
\end{equation}
$P(\beta,\gamma)$ is the shape distribution as a function of the
deformation coordinates $(\beta,\gamma)$ and $T$ is the temperature.  Plots are
truncated at small $\gamma$ for an obvious reason.  All shape plots
discussed in this paper are for the mass quadrupole and are not electric
quadrupole.

Moments of inertia are calculated in the cranked Hamiltonian
($\hat H\rightarrow \hat H-\omega \hat J_z$) using
${\cal I}_2=d\langle \hat J_z\rangle/d\omega$. 
We expect moments of inertia to initially increase as the nucleus is
cranked and then to decrease when pairs are broken.

\subsection{Pairing composition}
Pairing correlations in nuclei can be studied by calculating pairing
strengths in different spins for protons and neutrons.  For like
particle pairs, define the pair creation operator as

\begin{equation}
A^\dagger_{JM}=\frac{1}{\sqrt{1+\delta_{ab}}}[\hat a^\dagger_{j_a}\times
\hat a^\dagger_{j_b}]_{JM}(-)^{l_b}
\end{equation}
where
\begin{equation}
[\hat a^\dagger_{j_a}\times \hat a^\dagger_{j_b}]_{JM}=\sum_{m_a,m_b} (j_a m_a j_b
m_b|JM)\hat a^\dagger_{j_a}\hat a^\dagger_{j_b}
\end{equation}

Note that we only consider like particle pairing, i.e., proton-proton
and neutron-neutron (Eq.~\ref{eq:ham}).  Now let $\alpha=(j_a,j_b)$ and
$\alpha'=(j_c,j_d)$.  Using the pair 
creation operator, a matrix $M^J_{\alpha,\alpha'}$ can be constructed as 
\begin{equation}
\label{meq}
M^{J\pi}_{\alpha,\alpha'}=\sum_M \langle
A^\dagger_{JM}(j_a,j_b)A_{JM}(j_c,j_d)\rangle
\end{equation}
from which we can then define a pairing strength $P^J$ as
\begin{equation}
P^{J\pi}=\sum_{\alpha\ge\alpha'} M^{J\pi}_{\alpha,\alpha'}
\end{equation}
The correlated pair strength, which is more useful, is obtained by
subtracting uncorrelated mean field pairs from the total $P^{J\pi}$
defined above.  A Fermi gas has generally been used for the mean field
with SMMC.  Letting $n_k=\langle \hat a^\dagger_k \hat a_k\rangle$ and
substituting $n_1n_2(\delta_{13}\delta_{24}-\delta_{23}\delta_{14})$ for
$\langle \hat a^\dagger_1 \hat a^\dagger_2 a_3 a_4\rangle$ in
Eq.~\ref{meq} yields the Fermi gas mean field pair strength
$P^{J\pi}_{MF}$.  In this case, of course, we could use the SPA
occupations as the 
``mean field'' to subtract from the complete pairing plus quadrupole
solutions.

Even-even nuclei have correlated ground states, so we expect an excess
of $J~=~0$ pairs beyond the mean field in even-even ground states.  The
hallmark for a pair condensate is the existence of one eigenvalue of
$M^{J\pi}_{\alpha,\alpha'}$ which is much greater than all the rest.

The pair matrix can be diagonalized to find the eigenbosons
$B^{\dagger}_{\alpha JM\pi}$ as
\begin{equation}
B^\dagger_{\alpha JM\pi}=\sum_{a b} \psi_{\alpha JM\pi}(a b)
A^{\dagger}_{JM\pi}(a b)\;,
\end{equation}
where $\alpha=1,2,\cdots$ labels the
various bosons with the same angular momentum and parity. The
$\psi_{\alpha J\pi}$ are the eigenvectors of the diagonalization, i.e.
the wavefunctions of the boson, and satisfy the
relation
\begin{equation}
\sum_{j_aj_b}
\psi^*_{\alpha J\pi}\psi_{\mu J\pi} =\delta_{\alpha\mu}\;.
\end{equation}
These eigenbosons satisfy
\begin{eqnarray}\label{9}
\sum_M \langle B_{\alpha JM\pi}^\dagger  B_{\gamma JM\pi}\rangle =
n_{\alpha J\pi} \delta_{\alpha \gamma}\;,
\end{eqnarray}
where the positive eigenvalues $n_{\alpha J\pi}$ are  the number of
$J\pi$-pairs of type $\alpha$.

\subsection{Backbending}
We can monitor the pair strength for neutrons coupled to $J=12$ as a
signature for the anticipated band crossing or backbending.  The only
orbital in our model space which can produce this coupling is the
neutron $i_{13/2}$ level.  We do not monitor backbending by mapping
out the typical backbending plot of $I$ vs. $\omega$ because
the backbend in the plot requires a multivalued solution of $I$, whereas SMMC
always produces a single-valued solution from the statistical ensemble.

\subsection{Level density in SMMC}\label{rhoSMMC}
SMMC is an excellent way to calculate level densities.
$E(\beta)=\langle\hat H\rangle$ is
calculated for many values of $\beta$ which then determine the partition
function, $Z$, as
\begin{equation}
\ln[Z(\beta)/Z(0)]=-\int_0^\beta d\beta'E(\beta');
\end{equation}
$Z(0)$ is the total number of available states in the space.  The
level density is then computed as an inverse 
Laplace transform of $Z$.  Here, the last step is performed with a
saddle point approximation:
\begin{eqnarray}
S(E)& = & \beta E + \ln Z(\beta) \\
\rho(E)&=&(2\pi\beta^{-2}C)^{-1/2}\exp(S),
\label{eq:rho}
\end{eqnarray}
where $\beta^{-2}C\equiv -dE/d\beta$.  SMMC has been used recently to
calculate level densities in iron region nuclei~\cite{density}.  Here,
we present the first exact level density calculation for the much heavier Dy.

Nuclear level densities in the static path approximation has
previously been investigated by Alhassid and Bush for a 
simple solvable Lipkin model~\cite{spa1}.  The simple Lipkin
Hamiltonian does not include pairing, however.  These authors found SPA to be
superior to the mean field approximation and the difference between
the two depended on interaction strength.   

\subsection{Recap on interaction and model space}
\label{modelspace}
It is fortunate that the elementary pairing plus quadrupole
interaction does not break the Monte Carlo sign so that no
g-extrapolation is required.  The calculation of level densities is
also simplified, since accurate results require very low statistical 
uncertainties~\cite{density}.

We do not include isovector proton-neutron (pn) pairing in the
interaction.  This is a 
reasonable assumption since fp shell calculations with SMMC have shown clearly
that isovector pn correlations diminish quickly as N exceeds Z and
we are not at N$\sim$Z~\cite{fppair}.  Further, in our case the valence
protons and neutrons occupy different oscillator shells.  To be sure
some pn correlations are included via the isoscalar quadrupole interaction
($2\hat Q_p\cdot\hat Q_n$), but any observable that depends strongly on pn
correlations such as the Gamow-Teller strength will not be accurately
determined with this interaction. 

We choose one shell each for protons (sdg) and neutrons (pfh)
with the opposite parity intruders $h_{11/2}$ and $i_{13/2}$,
respectively.  The space encompasses $32$ proton
levels and $44$  
neutron levels.  The oscillator length, $b=\sqrt{\hbar/m\omega_0}$,
was taken to be  $1.01A^{1/6}$ fm.  

With this space, Dy has sixteen valence protons so that the proton shell is
half-filled.  The number of valence
neutrons varies from four in $^{152}$Dy to fourteen in $^{162}$Dy.  This model
space is identical to the one used by Kisslinger and 
Sorensen~\cite{Kisslinger-Sorensen}, but is smaller than the two shell
space Kumar and Baranger used. 

\subsection{Static path approximation and mean field}
\label{spavsmf}
The static path approximation (SPA) is the one-time-slice
limit of the partition function path integral and is obviously easily
implemented in SMMC.  The mean field
approximation is the saddle point estimation of the path integral;
neither includes imaginary time-dependent terms.  The SPA differs from mean
field since the integral over time-independent auxiliary fields in SPA
is done exactly, so that contributions from multiple configurations
(even with large fluctuations) are included rather than just the
single, steepest-descent mean field. 

Static path calculations do not do as well with realistic interactions
as with schematic interactions, e.g., pairing plus quadrupole.  Also,
the accuracy of the approximation varies among operators, as will be
demonstrated in Section~\ref{results}.

Static path calculations were done in SMMC with continuous fields rather than
the discrete field sampling that has typically been used with SMMC.
However, 
continuous sampling is not used for other calculations due to the
significantly increased number of thermalization cycles it would 
require.  Continuous fields proved necessary for accurate results in SPA;
calculated energies were higher for discretized fields than for
continuous fields and the difference could not be attributed to
statistical uncertainties.

\section{Results}
\label{results}
\subsection{SPA vs. full solution: static properties}
\subsubsection{Energy, spin, and quadrupole moments}
\label{static}
\begin{figure}
\begin{center}
\hskip 0.5in \psfig{figure=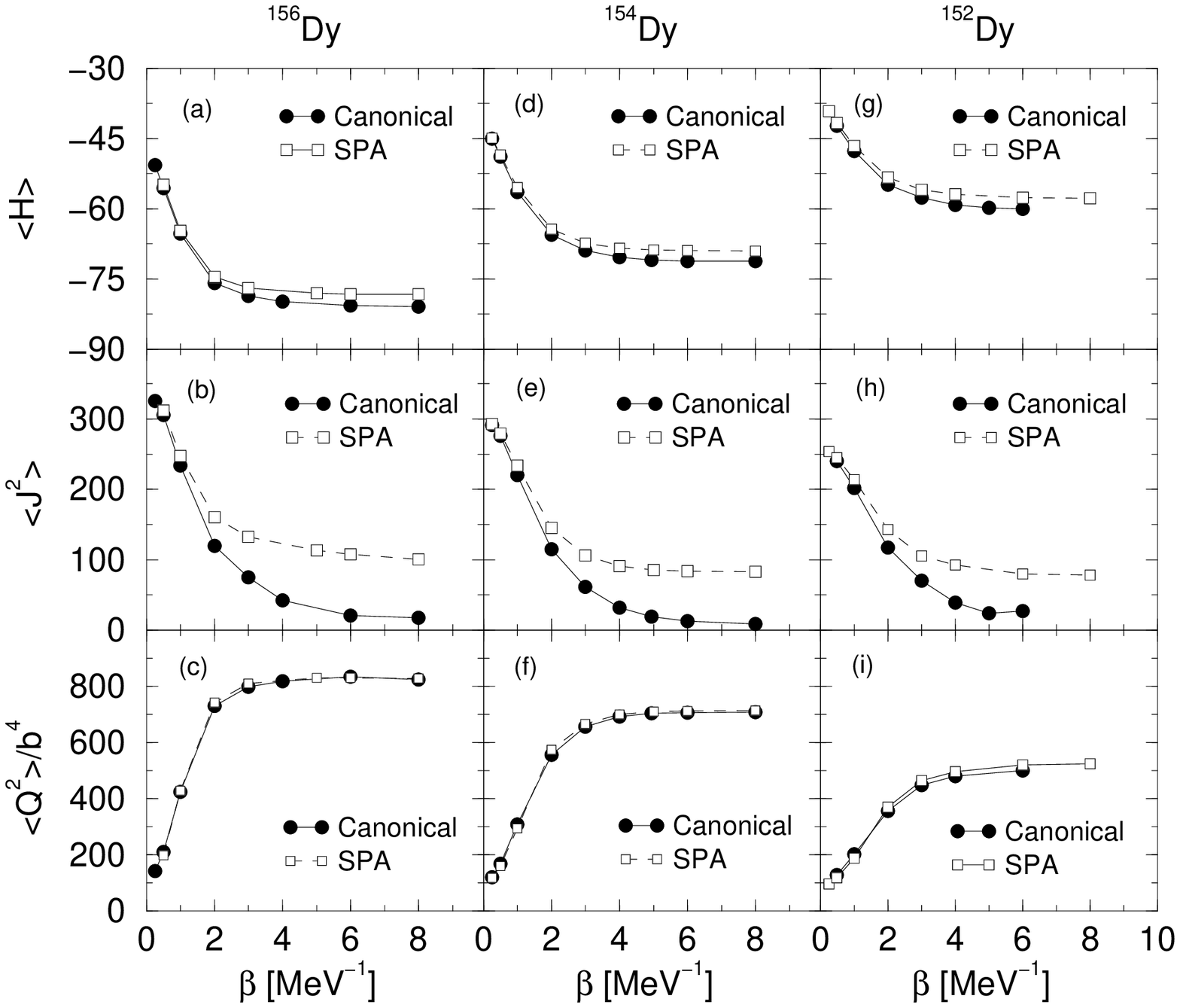,height=14cm}
\caption{(a)-(c) Energy, spin, and quadrupole moment for $^{156}$Dy.
(d)-(f) Energy, spin, and quadrupole moment for $^{154}$Dy.  (g)-(i)
Energy, spin, and quadrupole moment in  $^{152}$Dy.} 
\label{fig:static}
\end{center}
\end{figure}

Comparisons of SPA and full path solutions for the static
observables energy $\langle H\rangle$, spin $\langle
J^2\rangle$, and mass quadrupole moment $\langle Q^2\rangle$ are shown
for the experimentally deformed, transitional, and spherical isotopes in 
Figs.~\ref{fig:static}(a)-(i).  A few of the full canonical calculations
do not extend quite as low in temperature as the SPA
results due to numerical instability developing from multiple matrix
multiplications.  Error bars in these plots are smaller than the dot
sizes and are therefore not shown.  Also, note that the
quadrupole moments are expressed as $\langle Q^2\rangle/b^4$, where
$b$ is the oscillator length.

The SPA energy is greater than the exact energy, except at
very high temperatures for all three of these nuclei;  at the lowest
temperatures, the difference is a few MeV.  The origin of
this discrepancy will be discussed below (Section~\ref{pairenergy}).  
The difference between SPA and exact canonical ground state energies is
$2.39\pm0.15$ MeV for $^{152}$Dy and $2.59\pm0.18$ MeV for $^{156}$Dy,
so there is not a significant discrepancy between the lighter and
heavier isotopes.  Thus, SPA does not predict absolute energies 
accurately, but works well for relative differences.  The partition
function integral in SMMC is always divided into time steps of fixed
size $\Delta\beta$, which is fixed at 0.0625 MeV$^{-1}$.  As the
temperature, $1/\beta$, increases the number of time slices in the
exact partition function expression decreases.  Hence, it is not
surprising that SPA is more accurate for higher temperatures.

Looking at $\langle J^2\rangle$ shows that the SPA calculations only
cool to about $J=8-10$ for even the lowest calculated temperatures
while the even-even ground states are of course $J=0$.  In these
canonical SMMC calculations, $J=0$ is not exactly reached even in the
full canonical calculations because the
thermal ensemble always includes contributions from
higher energy states.  An estimate for the $\beta$ required for good
filtering to the even-even ground states is $\beta=1/E(2_1^+)$, where
$E(2_1^+)$ is the measured energy of the first $2^+$ state.
This $\beta$ value varies from $\beta=1.6$ MeV$^{-1}$  for $^{152}$Dy to
$\beta=11.5$ MeV$^{-1}$ for $^{162}$Dy as $E(2_1^+)$ varies from
0.614 to 0.087 MeV. 

The thermal spin
expectation, $\langle J^2\rangle$, can, in principle, be compared
against the experimental spectra.  However, one never experimentally knows all 
the states in a nucleus so such a direct comparison is difficult except at
low excitation energies which are dominated by the well known
ground state band.  The difference in $\langle J^2\rangle$ between SPA
and full canonical 
solutions at the lowest temperatures is $58$ for $^{152}$Dy, $74$ for
$^{154}$Dy, and $91$ for $^{156}$Dy.  Hence, the deviation is worse
with increasing deformation in these isotopes.

SPA works very well for the quadrupole moments.  This result is also very
robust, i.e., strengthening or weakening the coupling $\chi$ beyond the nominal
Kumar-Baranger value does affect the agreement between SPA
and exact results.

\begin{figure}
\begin{center}
\hskip 0.5in \psfig{figure=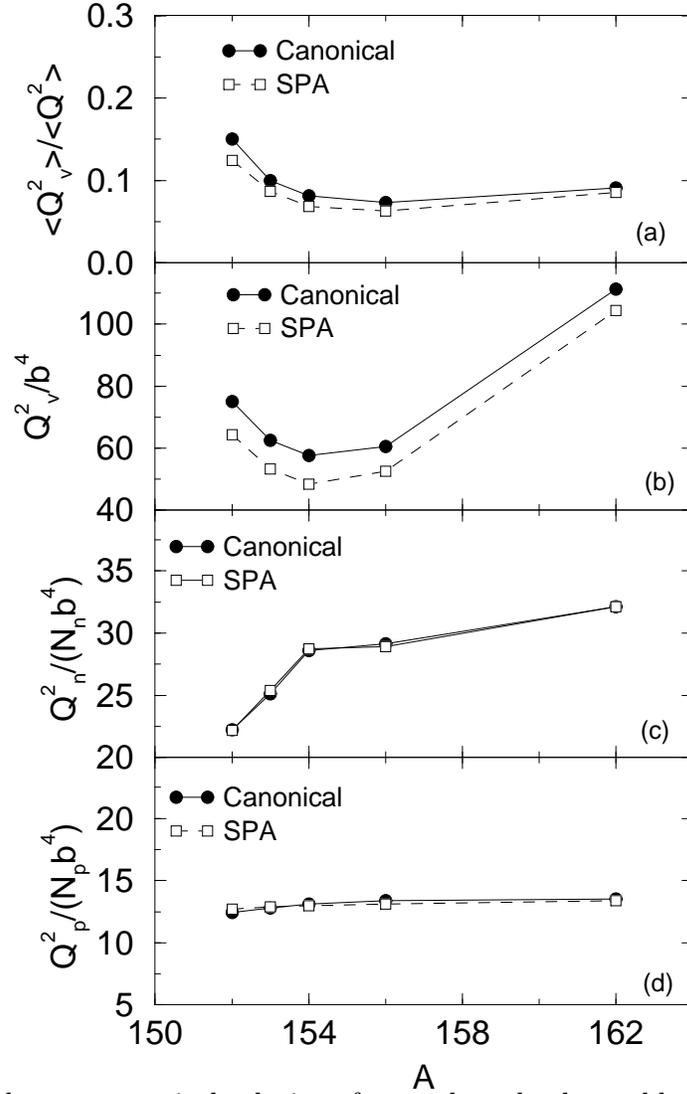,height=14cm}
\caption{SPA and exact canonical solutions for quadrupole observables in
selected dysprosium isotopes $A=152$ 
to $A=162$.  All results are for $\beta=6$ or $T=0.167$ MeV.  $\langle
Q_v^2\rangle$ is defined as $(Q_p-Q_v)^2$ and $b$ is the oscillator length.}
\label{fig:q}
\end{center}
\end{figure}

More information about the static quadrupole moment appears in
Figs.~\ref{fig:q}(a)-(d),
where the proton and neutron quadrupole components are shown
separately.  The neutron quadrupole moment increases quickly from
$^{152}$Dy to $^{154}$Dy and increases much more slowly after the onset of
deformation in $^{156}$Dy.  The proton quadrupole moment, meanwhile, 
remains approximately fixed for all the studied isotopes.  $\langle
Q^2_v\rangle$ increases rapidly as $A$
increases from $154$ to  
$162$.  Relative to 
$\langle Q^2 \rangle$ it is only 8.5\% (Fig.~\ref{fig:q}(a)), 
but the total quadrupole
moment at $\beta=6$ MeV$^{-1}$ is 46\% larger for $^{162}$Dy than for
$^{156}$Dy ($1221\; b^4$ vs. $834\; b^4$).  $\langle Q^2_v\rangle/\langle
Q^2\rangle$ is roughly 10\% for all isotopes, where $Q_v$ is the
isovector quadrupole operator ($Q_p-Q_n$).

Another comparison between SPA and full canonical calculations is how
quickly the solutions cool for various observables.  For example, in
$^{152}$Dy $\langle H\rangle$ appears to have stabilized by
$\beta=6$ MeV$^{-1}$ in both SPA and the full solution.  The spin $\langle
J^2\rangle$ and quadrupole moment $\langle Q^2\rangle$ also appear to
have minimized near $\beta=6$ MeV$^{-1}$ in the SPA and exact calculations.
Similar results are evident in the other nuclei, except for $\langle
J^2\rangle$ in $^{156}$Dy which is clearly decreasing still in SPA for
$\beta=8$ MeV$^{-1}$, the largest value for which the calculation
could be done. 

\subsubsection{Pairing energy and gaps}
\label{pairenergy}
\typeout{check 156 pairing energy and gaps; graphs not consistent?}
Some insight can be had by looking at the pairing energy and gaps.  The
quadrupole energy,$-0.5\chi\langle \hat Q\cdot \hat Q\rangle$ in SPA agrees
well with the full canonical solution, 
but the total energies in SPA shown above have clear deviations from
the exact canonical results.  The difference in $\langle H\rangle$ is
due to the pairing energy.  The accuracy of the SPA in the pairing interaction
depends strongly on the pairing strength and is naturally
better for weaker pairing.

\begin{figure}
\begin{center}
\hskip 0.5in \psfig{figure=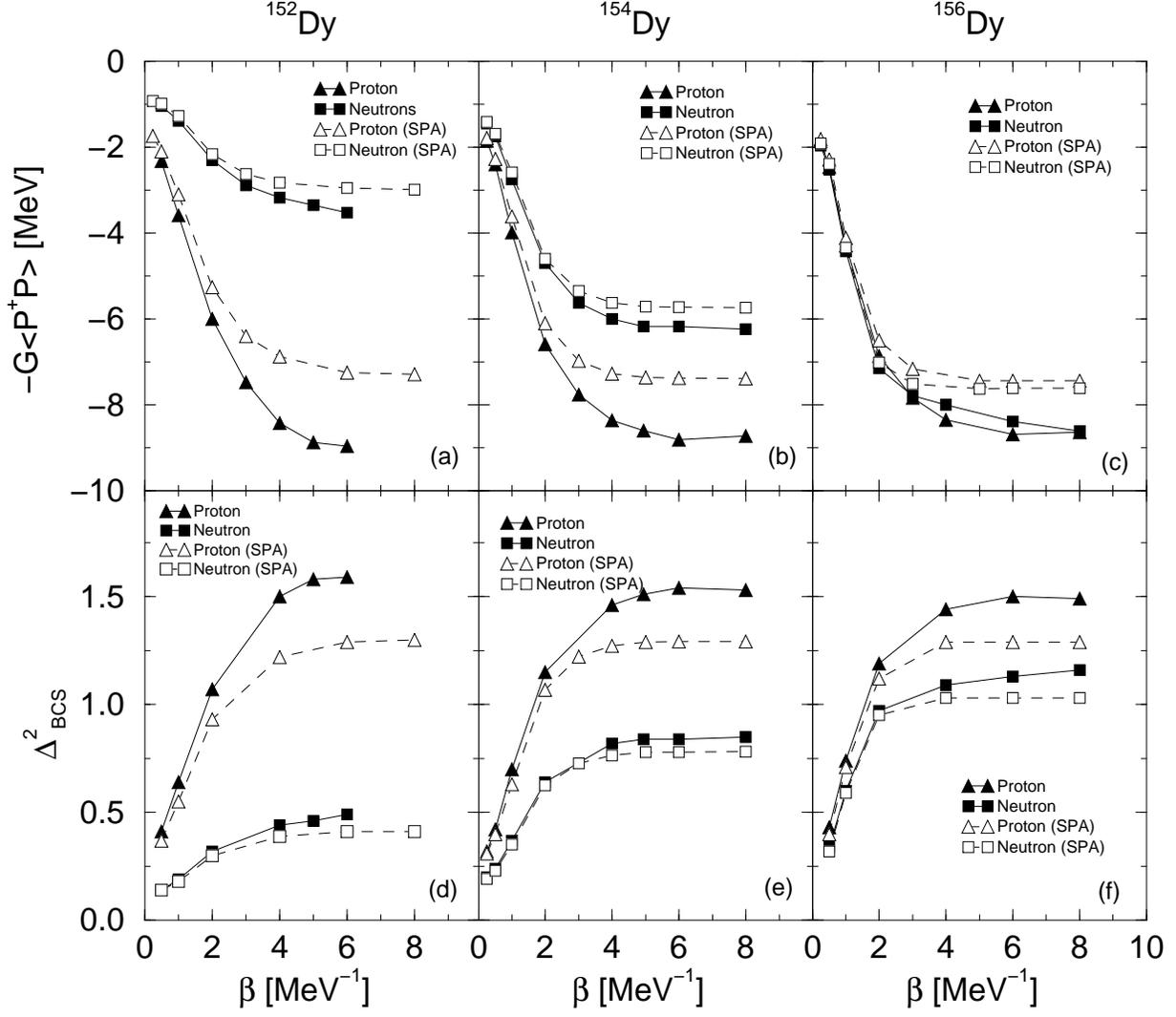,height=14cm}
\caption{Pairing energy in $^{152}$Dy (a), $^{154}$Dy (b), and
$^{156}$Dy (c).  BCS pair gaps are shown for $^{152}$Dy (d),
$^{154}$Dy (e), and $^{156}$Dy (f).}
\label{fig:tb}
\end{center}
\end{figure}

The pairing energies and BCS pair gaps for $^{152-156}$Dy are shown in
Figs.~\ref{fig:tb}(a)-(f).  The latter were obtained from
the pairing energies by $\Delta^2_{BCS}/G=H_{pair}$.  
The disagreement between SPA and exact
calculations looks worse for 
protons, but recall that there are 16 valence protons and just 4-8
valence neutrons in these isotopes. 

\subsection{B(E2) and effective charges}
\begin{figure}
\begin{center}
\hskip 0.5in \psfig{figure=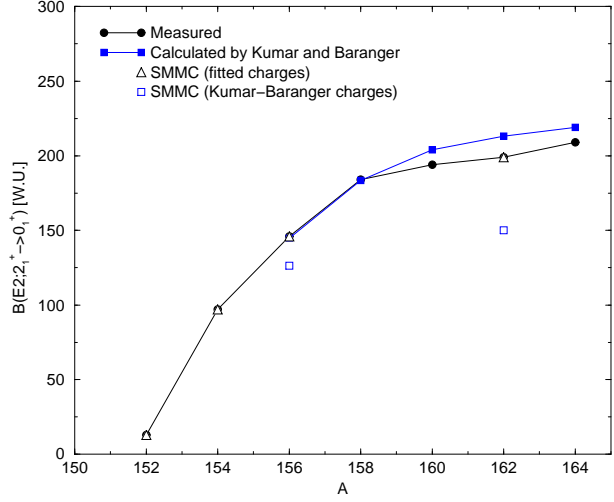,height=8cm,angle=-90}
\caption[$B(E2)$ in SMMC]{SMMC results for $B(E2;2_1^+\rightarrow 0_1^+)$ for
  dysprosium isotopes.  Results also shown in deformed cases (A=156,
  A=162) for strengths calculated with SMMC quadrupole moment and
  Kumar-Baranger effective charges.}
\label{fig:smmcbe2}
\end{center}
\end{figure}

The reduced electric quadrupole transition strength, $B(E2)$, is computed from 
\begin{equation}
B(E2)=\langle(e_p \hat Q_p+e_n \hat Q_n)^2\rangle
\end{equation}
where $e_p$ and $e_n$ are effective proton and neutron charges.  We
have taken $e_p=1+x, e_n=x$.  Results are shown 
in Table~\ref{table:be2} and Fig.~\ref{fig:smmcbe2}, where it has been
assumed that the total 
$B(E2)$ calculated in SMMC is the same as $B(E2;2_1^+\rightarrow
0_1^+)$.  Effective charges in column 3 and column 4 are fitted to
measured $B(E2)$ values.  Typical effective charges in rare earths are
approximately $e_{p(n)}=2(1)$, so these values are in a reasonable
range.

Table~\ref{table:be2} and Fig.~\ref{fig:smmcbe2} also show what
$B(E2)$ strength would be obtained from SMMC quadrupole moments with
Kumar-Baranger charges.  This illustrates how a small change in the
effective charge can produce a comparatively large change in the
$B(E2)$.  For $^{162}$Dy, the $7\%$ difference in $e_p$ when using
Kumar-Baranger charges leads to a $23\%$ change in $B(E2)$.

\begin{figure}
\begin{center}
\hskip 0.5in \psfig{figure=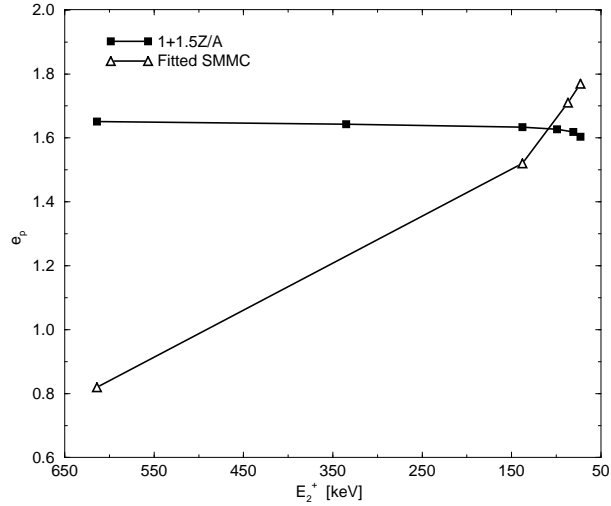,height=8cm,angle=-90}
\caption[Effective charge vs. $E(2_1^+)$ in Dy]{Effective charge
  vs. $E(2_1^+)$ in Dy.  Shown for Kumar-Baranger effective charge in
  deformed nuclei, $e_p=1+1.5Z/A$, and fitted SMMC charges.
  $E(2_1^+)=614$ keV is spherical $^{152}$Dy and $E(2_1^+)=335$ keV is
  $^{154}$Dy.  The other points are for deformed $^{156}$Dy and
  $^{162}$Dy.}
\label{fig:e2plus}
\end{center}
\end{figure}

The collectivity of a nucleus, and thus the $B(E2)$ value, varies with
the energy $E(2_1^+)$.  The effective proton charge, $e_p$, is plotted
against $E(2_1^+)$ in Fig.~\ref{fig:e2plus}.  For the SMMC results,
the neutron effective charge needs to be zero for spherical $^{152}$Dy to
avoid severely overestimating the $B(E2;2_1^+\rightarrow 0_1^+)$
strength as it is calculated above.  The fitted effective charge is
$e_p=1.75$ for the deformed isotopes and it is intermediate for
$^{154}$Dy.  This is a reflection of the fact that exact solutions for
the mean field interaction yield lighter dysprosiums which are too
deformed (Section~\ref{sts}); the effective charge should be constant.

It should be noted that Kumar and Baranger did not calculate $B(E2)$
values for spherical nuclei in the same way as for deformed nuclei,
i.e., they did not take effective charges for spherical nuclei as
$e_p=1+1.5Z/A$, $e_n=1.5Z/A$.  For spherical nuclei, they combined
phonon and rotational model properties (see~\cite{Kumar-Baranger}
p. 552).  They used $E(2_1^+)=(C/B)^{1/2}$ and
$B(E2,0\rightarrow2)\propto Z^2R_0^4 (BC)^{-1/2}$ from the phonon model  
with the relation $Q_s\propto B(E2)^{1/2}$.  These $E(2_1^+)$
predictions are not very good (Table~\ref{table:kbspherbe2}). 

\subsection{SPA vs. full solution: cranking}
\subsection{Sign limits on cranking}
\begin{figure}
\begin{center}
\hskip 0.5in \psfig{figure=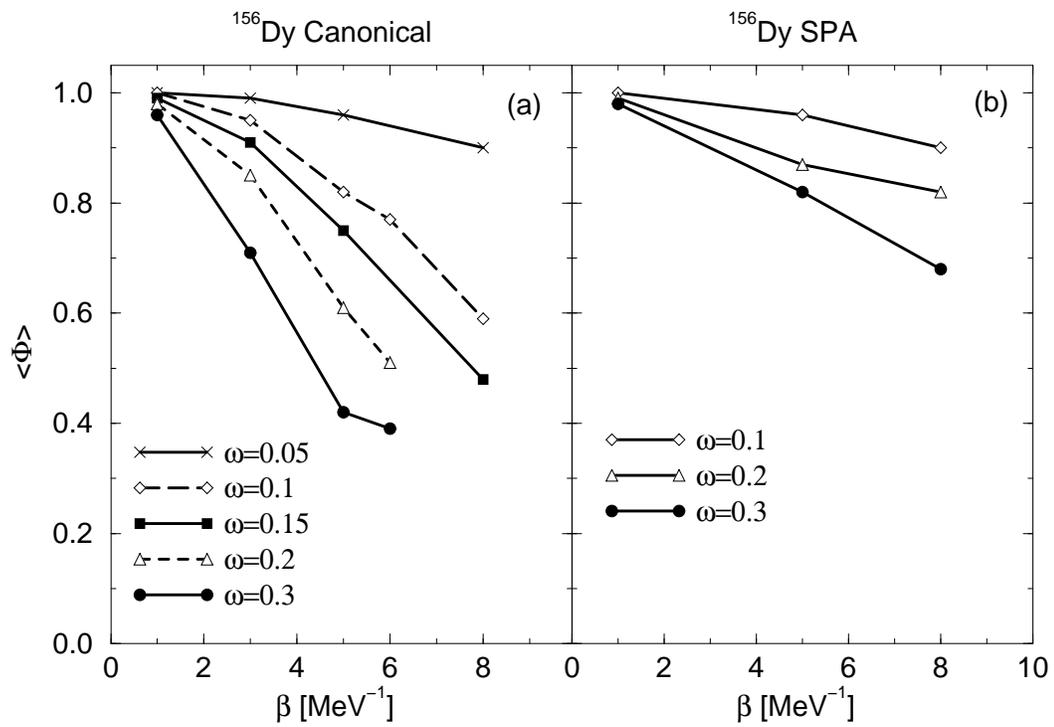,height=12cm}
\caption{Monte Carlo sign for canonical (a) and SPA (b) cranking in
$^{156}$Dy.} 
\label{fig:csign}
\end{center}
\end{figure}

Recall from Section~\ref{sign} that cranking degrades the Monte
Carlo sign from unity and that calculations become
impractical when the sign drops below 0.5.  This is illustrated for
both full canonical 
and SPA results for $^{156}$Dy in Figs.~\ref{fig:csign}(a)-(b).  Error
bars are not displayed in these figures since they are small; the
statistical error in $\Phi$ for $\beta=6$ MeV$^{-1}$ and $\omega=0.3$
MeV is 0.04.  Canonical cranking in this nucleus is limited to small
frequencies for 
$\beta=8$ MeV$^{-1}$, as evidenced by the quick drop in sign from $\omega=0.05$
MeV to $\omega=0.1$ MeV.  The canonical cranking is fairly good for
$\beta\leq6$ MeV$^{-1}$,
especially $\beta\leq4$ MeV$^{-1}$.  SPA cranking predictably has better sign
properties.  In SPA, $^{156}$Dy can be cranked well out to $\beta=8$
MeV$^{-1}$,
which is the approximate limit of temperature that can be reached in
this nucleus without matrix stabilization.

\subsubsection{Energy, spin, and quadrupole moments again}
\begin{figure}
\begin{center}
\hskip 0.5in \psfig{figure=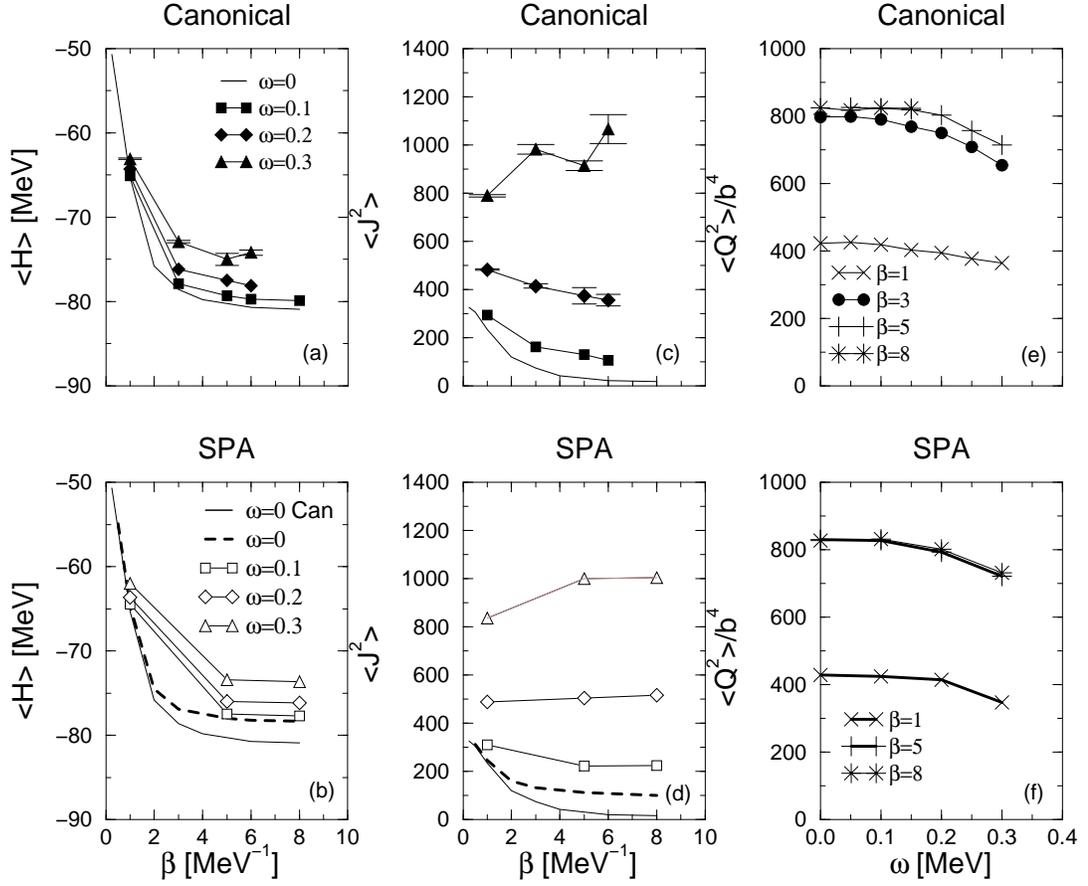,height=12cm}
\caption{Canonical vs. SPA
  cranking results in $^{156}$Dy for energy (a)-(b), spin (c)-(d), and
  quadrupole moment (e)-(f).  Error bars are not shown for SPA results
  since error bars are smaller than the symbols.}
\label{fig:scrank}
\end{center}
\end{figure}

Energy, spin, and quadrupole moments at various temperatures
are compared at different cranking frequencies in exact canonical
and SPA.  Calculations for $^{156}$Dy appear in
Figs.~\ref{fig:scrank}(a)-(f).  Energy results for
SPA at different cranking frequencies mirror the full canonical
result.  In the range $5\leq\beta\leq8$ MeV$^{-1}$, or $0.2\leq T\leq
0.125$ MeV, 
$\omega=0.1$ MeV lies at small 
excitation energy $\epsilon$ above the $\omega=0$ baseline,
$\omega=0.2$ MeV lies
roughly $3\epsilon$ above the baseline, and $\omega=0.3$ MeV is excited 
by approximately $6\epsilon$ above $\omega=0$.  

The spin results reveal that $\langle J^2\rangle$ in SPA is higher
than the exact canonical result until $T\geq1$ MeV when
$\omega\leq0.2$ MeV.  However, for $\omega=0.3$ MeV, SPA agrees well with the
exact solution.  The exact solution is not shown for $\beta=8$
MeV$^{-1}$ at this
frequency due to numerical difficulties.  The SPA $\langle J^2\rangle$
for $\omega=0.2$ MeV is very flat across the 
computed temperature range, $0.125\leq T\leq 1$ MeV.  As with the
exact solution, $\langle J^2\rangle$ decreases with
rising temperature for $\omega=0.3$ MeV.

Quadrupole results with cranking are similar to the $\omega=0$ results
in that the quadrupole moment does not change when the
temperature is decreased below 200 keV at any frequency studied here.
Note that 
the quadrupole results are plotted vs. $\omega$ for various
temperatures.  At the lowest temperatures, the quadrupole moment begins
to decrease after $\omega=0.15$ MeV in the canonical case.  It decreases
after $\omega=0.1$ MeV in the SPA, however, $\omega=0.15$ MeV is not computed
there so it is difficult to say if the quadrupole moment is
declining at frequency $0.1$ or $0.15$ MeV in SPA.  SPA agrees very well
with the exact solution for $\langle Q^2\rangle$ at all temperatures
and cranking frequencies computed.

\subsubsection{Moments of inertia}
\begin{figure}
\begin{center}
\hskip 0.5in \psfig{figure=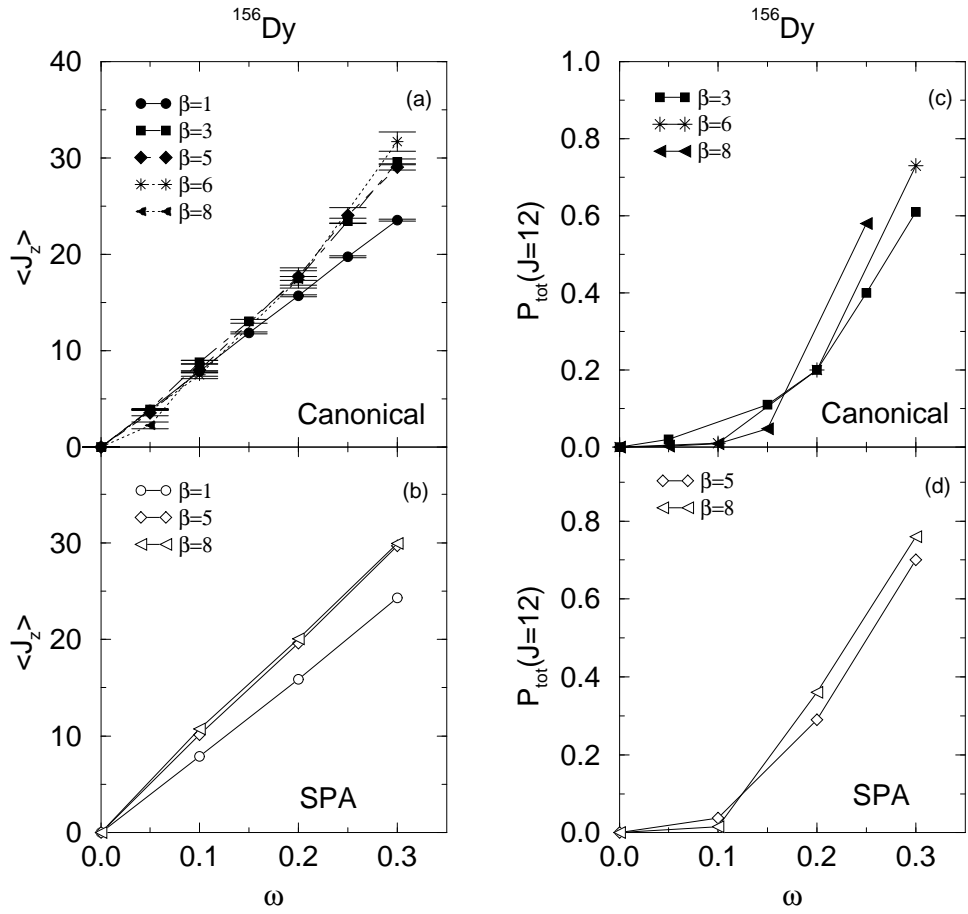,height=12cm}
\caption{(a)-(b) $\langle J_z\rangle$ for canonical and SPA cranking in
  $^{156}$Dy.  (c)-(d) $J=12$ pair strength in canonical and SPA
  cranking in $^{156}$Dy.  $J=12$ pair strength comes exclusively from
  $i_{13/2}$.}
\label{fig:jzp}
\end{center}
\end{figure}

The $J_z$ variation with cranking frequency determines the moment of
inertia.  Results for $^{156}$Dy are
displayed in Figs.~\ref{fig:jzp}$(a)-(b)$ for both canonical and
SPA cases.

The moment of inertia, ${\cal I}_2$, for $^{156}$Dy is $44.6\pm7$
$\hbar^2$/MeV in the exact canonical ensemble at $J\approx 4$ with
$\beta=8$ MeV$^{-1}$ and is
$73.0\pm2$ $\hbar^2$/MeV at same $\omega$ in SPA.  At this temperature,
$T=0.125$ MeV, the SPA moment 
of inertia is $64\%$ larger than the exact result.
The experimental moment of inertia ${\cal I}_2=40 \hbar^2$/MeV
at $J=4$, which matches the SMMC canonical result.  Also, the
rigid body moment of inertia for $^{156}$Dy with 
$(\beta,\gamma)=(0.24,0)$ is $73 \hbar^2$/MeV.  This coincides with
the SPA moment of inertia.

\subsection{Band crossing}

The pairing strength for $J=12$ pairs in $^{156}$Dy, which can be
produced only from $i_{13/2}$ neutron pairs, is shown in
Fig.~\ref{fig:jzp}(c)-(d).  The strength $P_{J=12}$ begins to
increase quickly 
in the canonical case for $(\beta,\omega)=(8,0.15)$, which corresponds
to $J=16\pm1$.  In SPA, $P_{J=12}$ increases sharply beyond
$(\beta,\omega)=(8,0.1)$, which corresponds to $J=14$.  

\begin{figure}
\begin{center}
\hskip 0.5in \psfig{figure=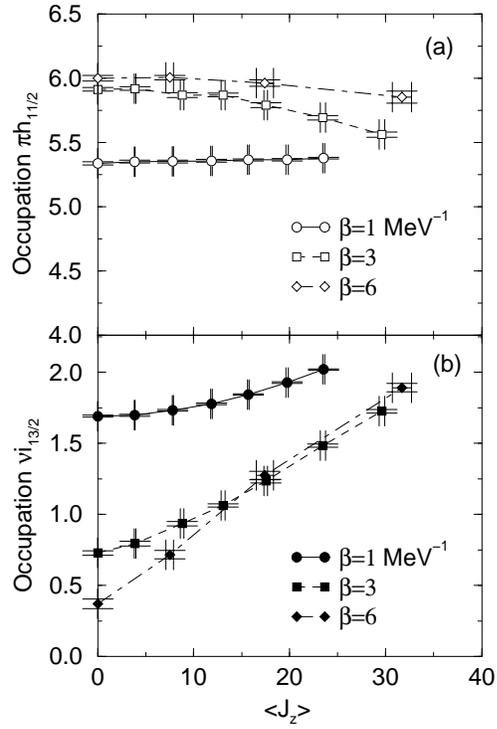,height=10cm}
\caption{Occupation for proton $h_{11/2}$ (a) and neutron $i_{13/2}$
(b) vs. spin.}
\label{intruderocc1}
\end{center}
\end{figure}

The occupations of both the proton $h_{11/2}$ and neutron $i_{13/2}$
intruder orbitals for $^{156}$Dy are given in 
Figs.~\ref{intruderocc1}(a)-(b).  The proton intruder
occupation is comparatively stable over this same spin range at each 
temperature.  However, it is clear that the $i_{13/2}$ occupation is
increasing with spin, particularly for lower temperatures, as
expected.  The $h_{11/2}$ occupation number 
decreases slightly with temperature for all frequencies computed.
Occupation shifts slightly to $\pi d_{5/2}$.
For $\beta=6$ MeV$^{-1}$, the maximum spin $J_z$ corresponds to
$J\approx32$ and for 
$\beta=1$ MeV$^{-1}$, the maximum spin is $J\approx28$.  Unfortunately for
$\beta=6$ MeV$^{-1}$, or $T=167$ keV, the Monte Carlo sign is reduced to $0.4$
at the maximum spin shown (recall Fig.~\ref{fig:csign}).  At
$\beta=1$ MeV$^{-1}$, however, the sign is still very stable at $0.96$.  

\subsection{Shape vs. temperature and spin}
\label{sts}
\begin{figure}
\begin{center}
\hskip 0.5in {\large (a) $^{152}$Dy, T$=0.25$ MeV} 
\hskip 0.5in {\large (b) $^{152}$Dy, T$=0.5$ MeV} \linebreak
\hskip 0.5in \psfig{figure=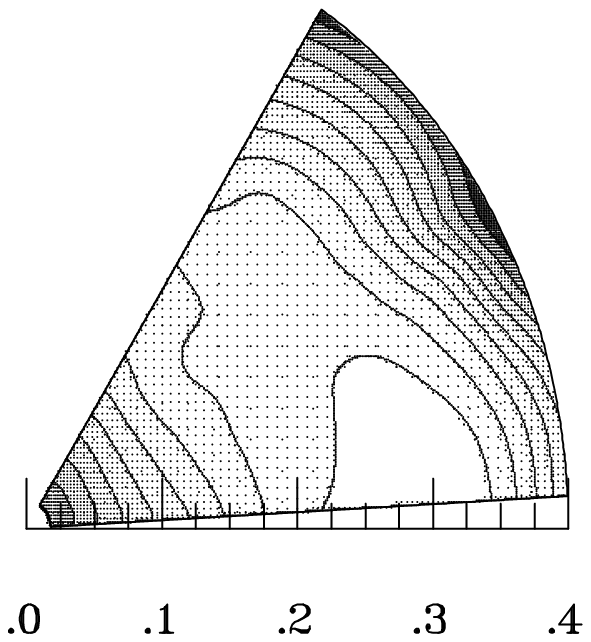,height=5cm} 
\hskip 0.5in \psfig{figure=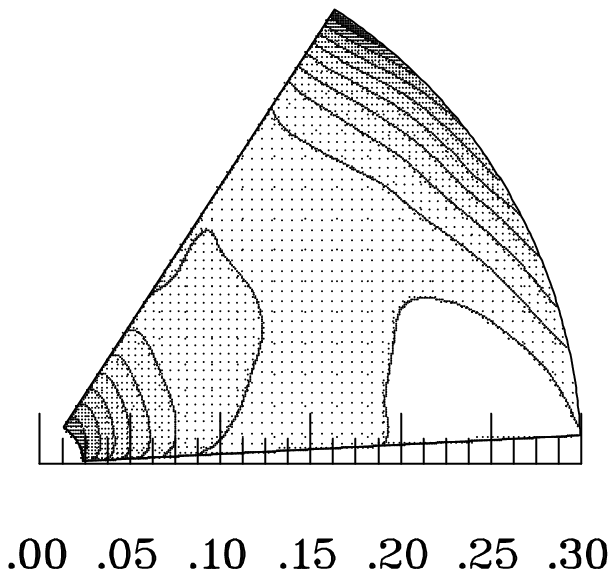,height=5cm} \linebreak

\hskip 0.5in {\large (c) $^{152}$Dy, T$=2$ MeV} \linebreak 
\hskip 0.5in \psfig{figure=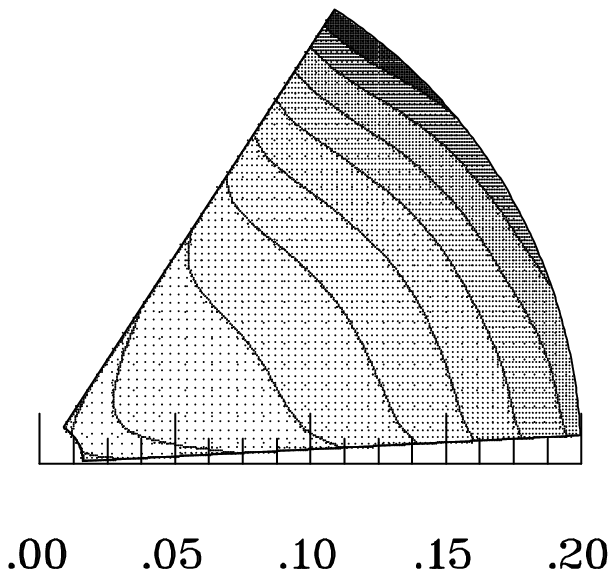,height=5cm} \linebreak
\caption{(a) Shape at T=0.25 MeV for
$^{152}$Dy. Contour spacing is 0.79 MeV. (b)
Shape at T=0.5 MeV for $^{152}$Dy.  Contour spacing is
0.64 MeV.  (c) Shape at T=2 MeV for $^{152}$Dy.  Contour
spacing is 2.9 MeV.  Each plot is compiled from 2000 sample.}
\label{fig:152shape}
\end{center}
\end{figure}

Nuclear shapes have also been computed to clarify how the shape varies
with temperature and spin.  Temperatures and frequencies for these
calculations are given in the figure captions.  In all shape graphs,
the $\beta$-axis is radial and the other axis is the $\gamma$-axis.
Results for $^{152}$Dy at temperatures  
from $T=0.25$ MeV to $T=2$ MeV are shown in
Fig.~\ref{fig:152shape}.  The nucleus becomes
increasingly spherical for rising temperature.  

\begin{figure}
\hskip 3.75in {\large $^{154}$Dy at T$=0.133$ MeV}
\begin{center}
\hskip 0.5in \psfig{figure=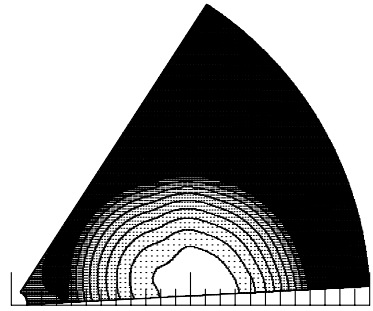,height=5cm}
\caption[Shape at T=0.133 MeV for $^{154}$Dy]{Shape at T=0.133 MeV for
$^{154}$Dy.  4800 samples.  Contour spacing is 0.92 MeV.}
\label{fig:154b7.5}
\end{center}
\end{figure}

\begin{figure}
\hskip 3in {\large $^{154}$Dy}
\begin{center}
\hskip 0.25in\psfig{figure=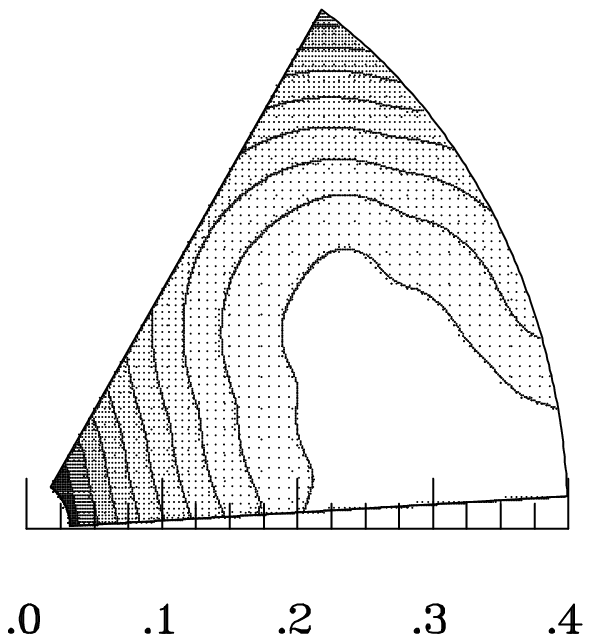,height=5cm}
\hskip 1.5in \psfig{figure=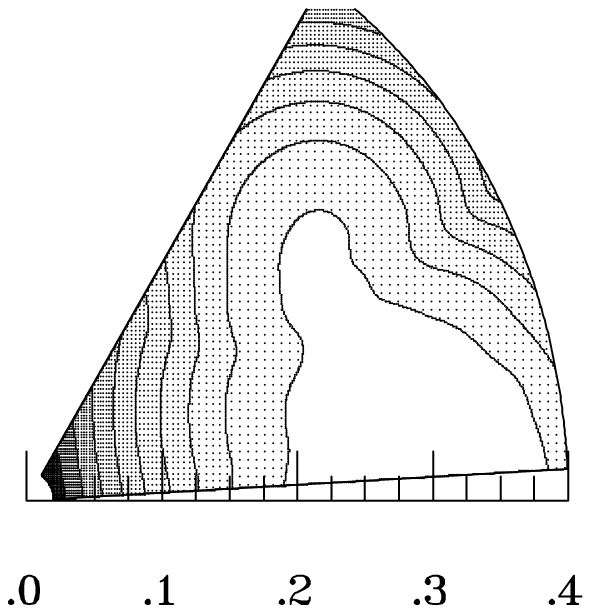,height=5cm} \linebreak
\quad (a) T$=0.25$ MeV, $\omega=0.05$ MeV$^{-1}$ 
\quad (b) T$=0.25$ MeV, $\omega=0.1$ MeV$^{-1}$ \linebreak
\hskip 0.25in \psfig{figure=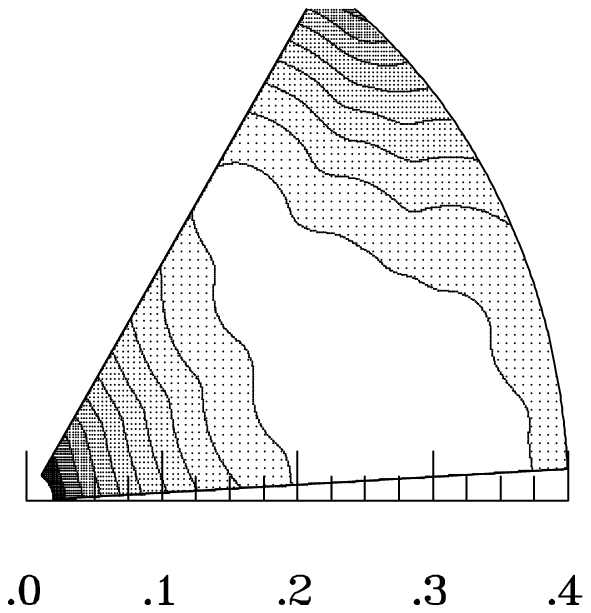,height=5cm}
\hskip 1.5in \psfig{figure=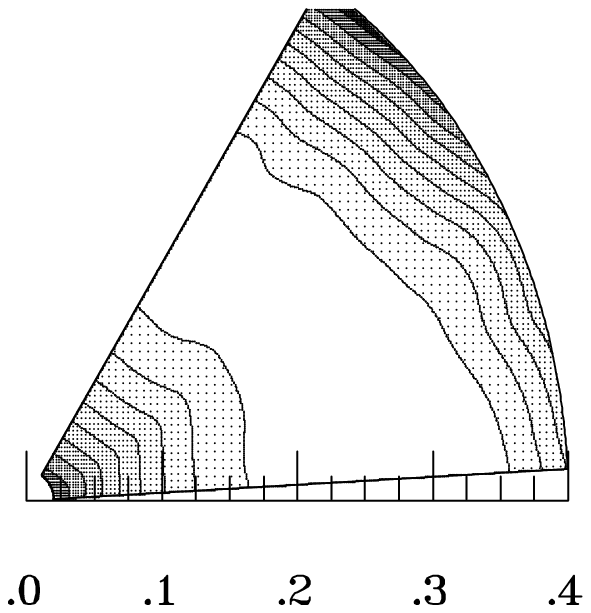,height=5cm} \linebreak
\quad (c) T$=0.25$ MeV, $\omega=0.2$ MeV$^{-1}$ 
\quad (d) T$=0.5$ MeV, $\omega=0.05$ MeV$^{-1}$ \linebreak
\hskip 0.25in \psfig{figure=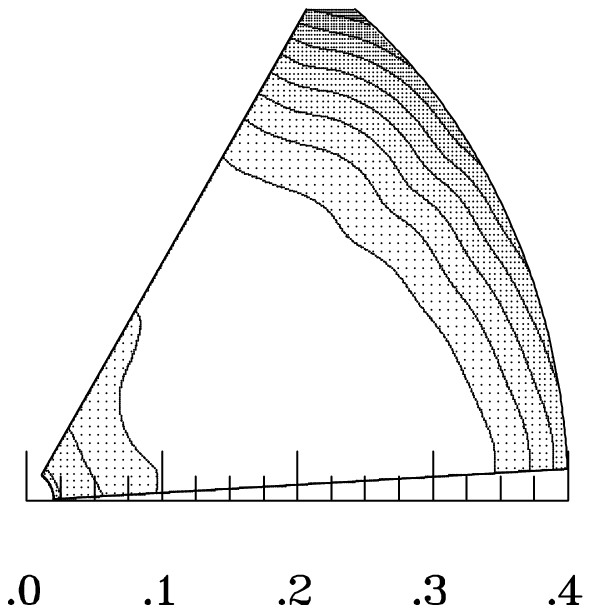,height=5cm}
\hskip 1.5in \psfig{figure=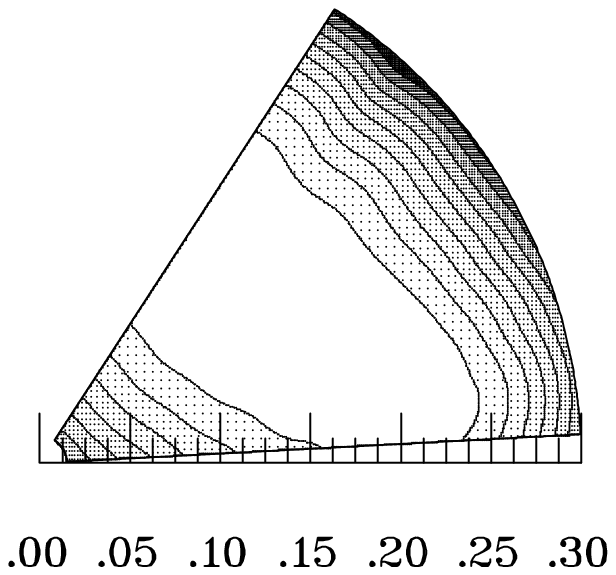,height=5cm} \linebreak
\quad (e) T$=0.5$ MeV, $\omega=0.2$ MeV$^{-1}$ 
\quad (f) SPA: T$=0.125$ MeV, $\omega=0.6$ MeV$^{-1}$ \linebreak

\caption{(a) Shape at T=0.25 MeV and $\omega=0.05$ ($J\approx6$) for  
$^{154}$Dy.  2560 samples.  Contour spacing is 2.9 MeV.
(b) Shape at T=0.25 MeV and $\omega=0.1$ ($J\approx 8$) for  
$^{154}$Dy.  2400 samples.  Contour spacing 2.5 MeV.
(c) Shape at T=0.25 MeV and $\omega=0.2$ for $^{154}$Dy.  2400
samples.  Contour spacing 2.4 MeV.
(d) Shape at T=0.5 MeV and $\omega=0.05$ ($J\approx 10$) for  
$^{154}$Dy.  3840 samples.  Contour spacing 3.4 MeV.
(e) Shape at T=0.5 MeV and $\omega=0.2$ ($J\approx 20$) for  
$^{154}$Dy.  1920 samples.  Contour spacing 4.2 MeV.
(f) Shape at T=0.125 MeV and $\omega=0.6$ ($J\approx 50)$ for
$^{154}$Dy in SPA.  Sign $\Phi=0.5$.  2000 samples.  Contour spacing
1.2 MeV.}
\label{fig:154shape}
\end{center}
\end{figure}

\begin{figure}
\hskip 2.75in {\large $^{156}$Dy} \linebreak
\begin{center}
\hskip -0.8in (a) T$=0.25$ MeV
\hskip 1.45in (b) T$=0.5$ MeV \linebreak
\hskip 0.5in \psfig{figure=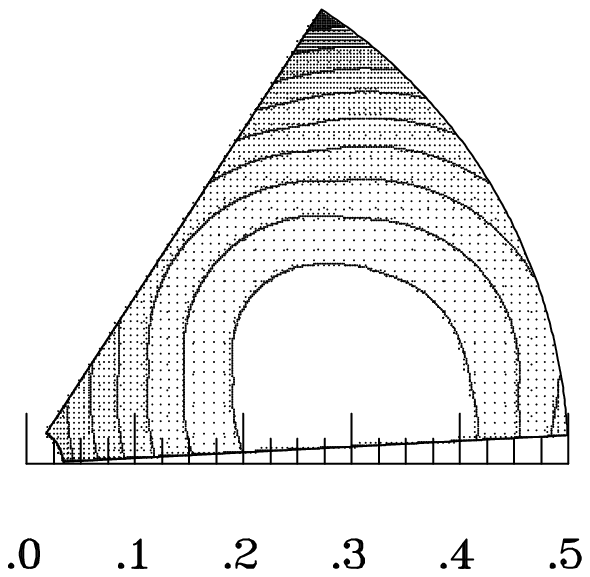,height=5cm}
\hskip 0.5in \psfig{figure=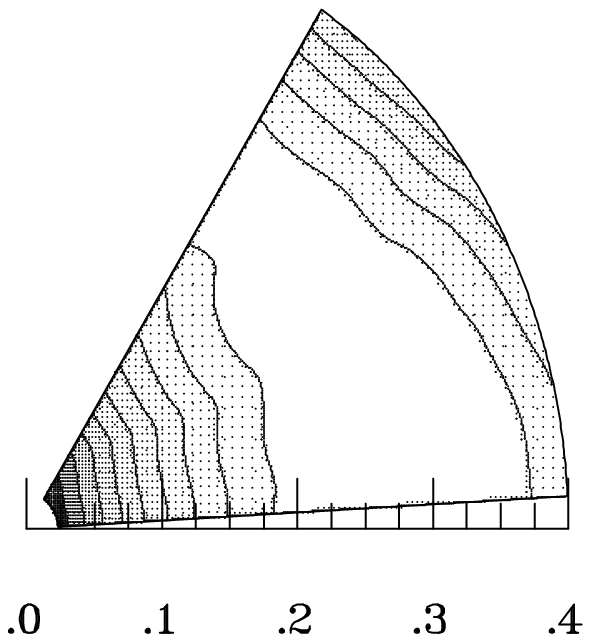,height=5cm} \linebreak
\hskip 0.75in (c) T$=1$ MeV
\hskip 1.6in (d) T$=0.125$ MeV, $\omega=0.1$ MeV$^{-1}$ \linebreak
\hskip 0.5in \psfig{figure=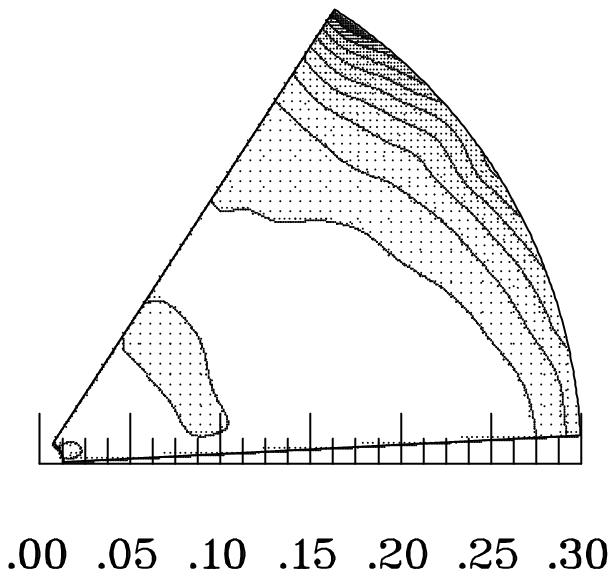,height=5cm}
\hskip 0.5in \psfig{figure=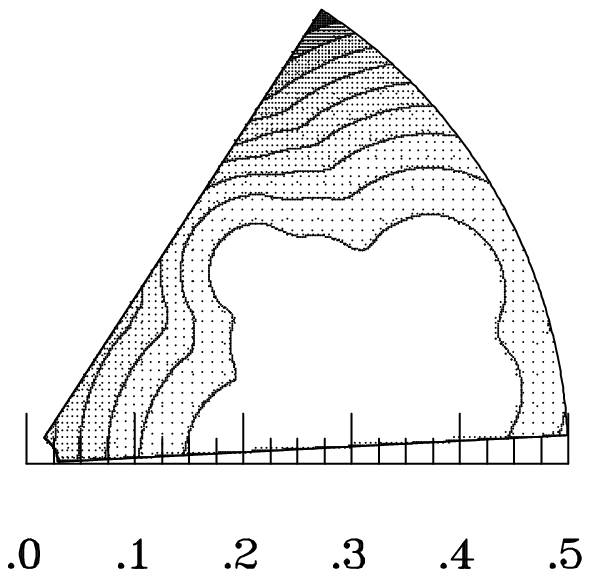,height=5cm} \linebreak
\caption{Shape at T=0.25 MeV for
$^{156}$Dy. 2000 samples.  Contour spacing is 6.5 MeV.
Shape at T=0.5 MeV for $^{156}$Dy. 2000 samples.  Contour spacing is
6.9 MeV.  
Shape at T=1 MeV for
$^{156}$Dy. 2000 samples.  Contour spacing is 2.0 MeV.
Shape at T=0.125 MeV for $^{156}$Dy with
$\omega=0.1$ MeV$^{-1}$. 4800 samples.  Contour spacing is 3.0 MeV.} 
\label{fig:156shape}
\end{center}
\end{figure}

Shapes for $^{154}$ Dy and $^{156}$Dy are shown in
Figs.~\ref{fig:154b7.5}-\ref{fig:156shape}.  These were all produced from
the exact canonical ensemble except Fig.~\ref{fig:154shape}(f), which
was produced with the static path approximation.  Cranked contour
plots, such as Fig.~\ref{fig:156shape}(d) for $^{156}$Dy at
$(\beta,\omega)=(8,0.1)$, show the nuclei becoming increasingly
gamma-soft with increasing spin.  This is also true in SPA
(Fig.~\ref{fig:154shape}(f)), which was utilized in
this case since the Monte Carlo sign for the exact calculation becomes
too small to obtain useful results.  There is no sign of oblate shape
at this spin in $^{154}$Dy, as predicted by Cranmer-Gordon, et al.,
using a Nilsson-Strutinsky cranking model~\cite{cranmer-gordon}.
However, the SMMC   
ground state deformation in $^{154}$Dy with these parameters is
clearly too large.  Ma, et al.~\cite{khoo154} claimed evidence for a
return to some collectivity in $^{154}$Dy from spin 36$^+$-$48^+$.  The
shape plot for $^{154}$Dy at $J\sim 50$ (Fig.~\ref{fig:154shape}(f))
appears soft.  The SMMC $B(E2)=5$ W.U. at this spin using the fitted
effective charge.  

Note that with increasing A in these isotopes, the ground state
deformation is roughly constant and the depth of the minimum
increases.  In fact for $^{156}$Dy, the depth of the well is roughly the same
as the fission barrier ($\sim40$ MeV)~\cite{Coulomb}.  However, it can
be questioned whether our shape formalism (Eq.~\ref{shapeeq}) should
be applied at such low temperatures ($T\leq 0.25$ MeV) since in the
limit $T\rightarrow 0$ the free energy becomes everywhere degenerate
and zero.  The very low temperature results for $^{156}$Dy and
$^{162}$Dy (not shown) do not coincide with the mean field.  

Previous publication of shape plots from SMMC results in gamma-soft
nuclei using a pairing plus quadrupole Hamiltonian with quadrupole
pairing~\cite{gamma-soft} 
did not exhibit this.  However, those nuclei are only weakly deformed
($\beta\approx 0.05-0.15$) while
the Dysprosiums with A$\geq$154 are well-deformed
($\beta\approx0.3$).  For the Dysprosium
shapes in this paper, the unexpected depth of the potential well is
only evident for well-deformed cases.

\begin{figure}
\hskip 2.25in {\large $^{140}$Ba at T$=0.25$ MeV}
\begin{center}
\hskip 0.5in \psfig{figure=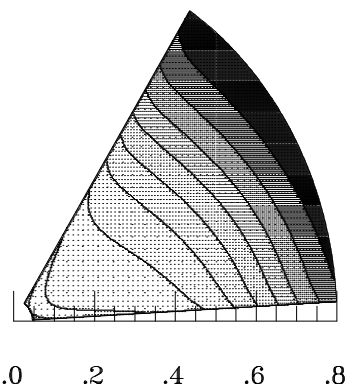,height=5cm}
\caption[Shape at $T=0.25$ MeV for $^{140}$Ba]{Shape at T=0.25 MeV for
$^{140}$Ba. 4000 samples.  Contour spacing 0.42 MeV.} 
\label{fig:ba140b4landdens}
\end{center}
\end{figure}

It is apparent from the above that the dysprosiums are all deformed in
their ground states in the exact model calculation.  However,
$^{152}$Dy is known experimentally to be spherical.  Kumar and
Baranger did not calculate $^{152}$Dy, though they did calculate some
other spherical isotopes.  A shape plot has
also been constructed from SMMC results in $^{140}$Ba for comparison
with Kumar and Baranger (Fig.~\ref{fig:ba140b4landdens}).  The SMMC
result for $^{140}$Ba agrees with Kumar-Baranger; both calculations indeed 
show a spherical nucleus.  This isotope
has $Z=56$ and $N=84$.  For the shell model space used, this becomes six
valence protons and two valence neutrons so that, unlike dysprosium,
the proton shell is less than half filled.  
The SMMC $B(E2;2\rightarrow 0)$ is $5$
W.U. using the effective charge fitted for the dysprosiums.
Reducing the quadrupole coupling to half its mean field strength still
yielded deformation $\beta\approx0.3$ for $^{156}$Dy with a deep
potential.  However, reducing $\chi$ to half the mean field strength
returns $^{152}$Dy to a spherical distribution which fits the measured
$B(E2)$ strength with effective charges $(e_p,e_n)=(1,0)$.

\begin{figure}
\hskip 2.25in {\large $^{144}$Ba at T$=0.25$ MeV}
\begin{center}
\hskip 0.5in \psfig{figure=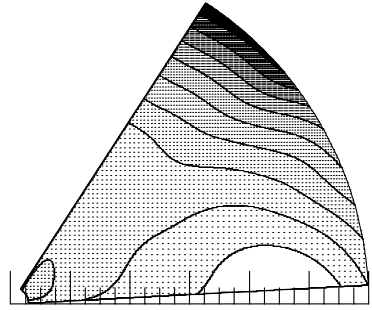,height=5cm}
\caption{Shape at $T=0.25$ MeV for $^{144}$Ba.  2300 samples.  Contour
spacing is 0.7 MeV.}
\label{fig:ba144b4bg}
\end{center}
\end{figure}

The equilibrium shape was also calculated in $^{144}$Ba for inverse
temperature $\beta=4$ MeV$^{-1}$. $^{144}$Ba has 
$E(2_1^+)=199$ keV~\cite{Isotopetable} and deformation
$\beta=0.19$~\cite{Raman}.  This 
nucleus proved to be extremely deformed  
($\beta\approx 0.45$) in SMMC 
with the Kumar-Baranger interaction parameters, but with a
deformation well not nearly so deep as for the $A\geq154$ Dysprosiums
(Fig.~\ref{fig:ba144b4bg}).  In this case, the potential was only about
2.5 MeV deep.  Kumar and Baranger did not calculate this isotope, so
direct comparison with them is not possible in this case.  Kumar and
Baranger also made no claims their model is valid for nuclei at such
extreme deformation~\cite{kb2}.

\subsection{Odd A}
\label{oddmass}
As mentioned previously, the odd nucleon in an odd mass nucleus
violates T-reversal symmetry 
and can break the Monte Carlo sign, even with an interaction free from
repulsive contributions.  Results from $^{153}$Dy are shown below in
Figs.~\ref{fig:153static}(a)-(e).  In 
this case, for our simple Hamiltonian the Monte Carlo sign behaves
well and remains at 0.82 for canonical $\beta=10$.  
Also, some of this reduction in sign may in fact be 
due to limits of numerical accuracy in the machine. 
\subsubsection{Static observables} 
\begin{figure}
\begin{center}
\hskip -1.1in \psfig{figure=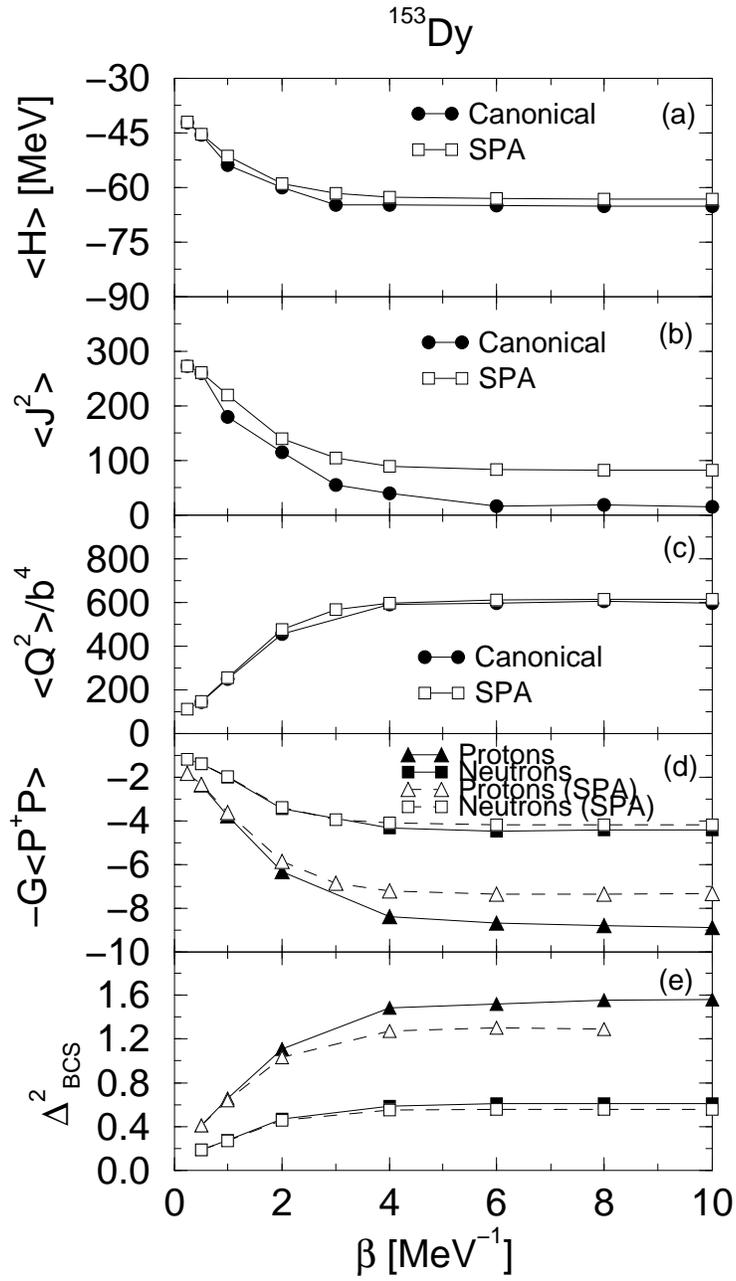,height=16cm}
\vspace{0.3in}
\caption{Energy (a), spin (b), quadrupole moment (c), pairing energy
(d), and BCS pair 
gap (e) in $^{153}$Dy for Kumar-Baranger interaction strengths.  Symbols 
for $(e)$ match the symbols in $(d)$.}
\label{fig:153static}
\end{center}
\end{figure}

Results for static observables are quantitatively similar to the
even-even results.  The energy difference 
between the full canonical and SPA calculations  
at $T=100$ keV is $1.97\pm0.4$ MeV (Fig.~\ref{fig:153static}(a)), which is
a little less than the $2.49\pm 0.16$ MeV canonical-SPA energy difference
in neighboring $^{152}$Dy and the $2.15\pm 0.06$ difference in
$^{154}$Dy.  This difference is due to different 
pairing energies in these odd-even and even-even isotopes.  The
discrepancy in $\langle J^2\rangle$ between SPA and full solutions 
is $\Delta J^2=66$ or $\Delta J\approx 8$ (Fig.~\ref{fig:153static}(b)).  The
ground state spin for $^{153}$Dy is $(7/2)^-$ so that $\langle
J^2\rangle=15.75$ and the first excited
state is $(3/2)^-$ at excitation $E=109$ keV.  Thus, the estimated
$\beta$ needed for filtering the ground state is reachable and
$\langle J^2 \rangle =16\pm4$ in the SMMC canonical
ensemble agrees very well with experiment.  Again, the
SPA quadrupole moment in 
$^{153}$Dy is in excellent agreement with the full canonical
calculation (Fig.~\ref{fig:153static}(c)).  The total pairing energy
and BCS gaps (Fig.~\ref{fig:153static}(d)-(e)) are similar to results in
$^{152}$Dy and $^{154}$Dy (Fig.~\ref{fig:tb}(a)-(e)).

\begin{figure}
\begin{center}
\hskip 0.5in \psfig{figure=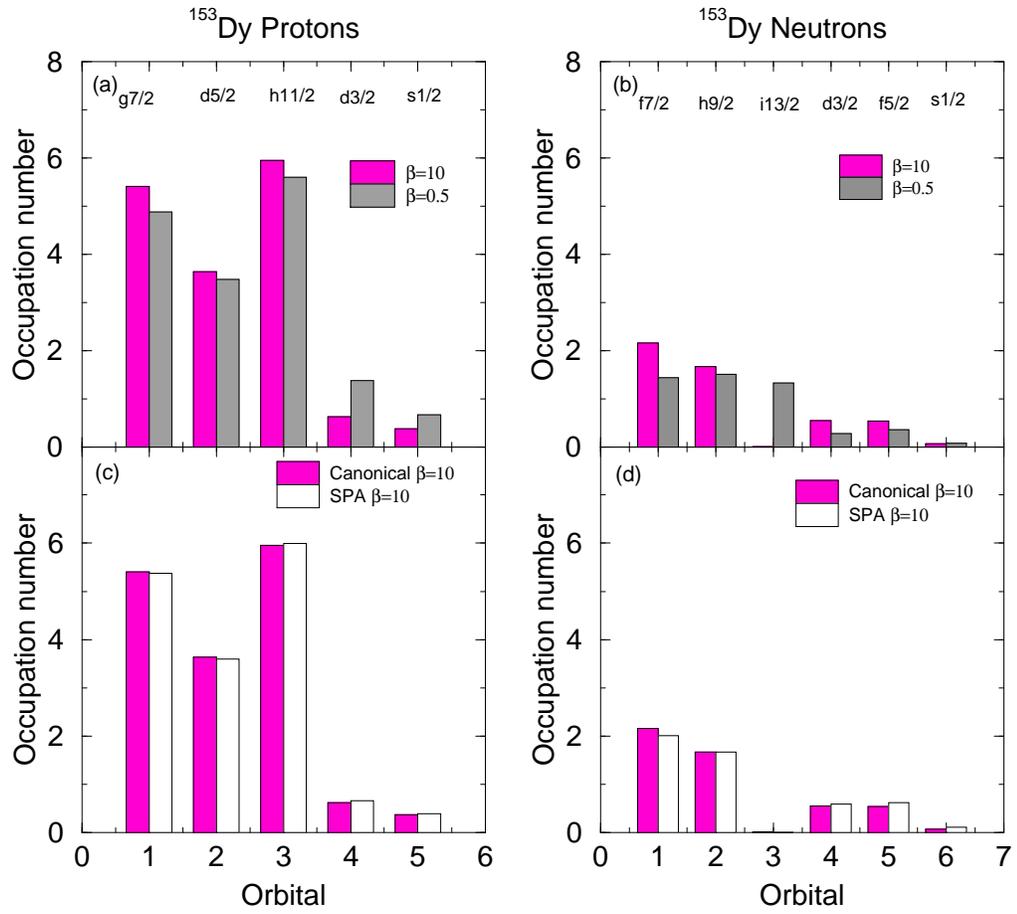,height=12cm}
\caption{Proton and neutron 
occupations in $^{153}$Dy.  Results (a)
and (b) are for the canonical ensemble at $T=0.1$ MeV and $T=2$ MeV
while (c) and (d) are canonical vs. SPA at $T=0.1$ MeV.} 
\label{fig:153occ}
\end{center}
\end{figure}

Occupation numbers for protons and neutrons in the canonical ensemble
for $^{153}$Dy appear in Fig.~\ref{fig:153occ}(a)-(b).  As the temperature
increases to $T=2$ MeV, the proton occupation shifts only slightly to
the highest orbitals.  For neutrons, however, there is a
clear rise in the $i_{13/2}$ occupation.  The occupation numbers are
also compared for full canonical vs. SPA in  
$^{153}$Dy for $\beta=10$ in Figs.~\ref{fig:153occ}(c)-(d). These
occupation numbers look very similar, 
though the agreement is slightly better for the protons.  Pairing
strengths are more revealing.

\begin{figure}
\begin{center}
\hskip 0.25in \psfig{figure=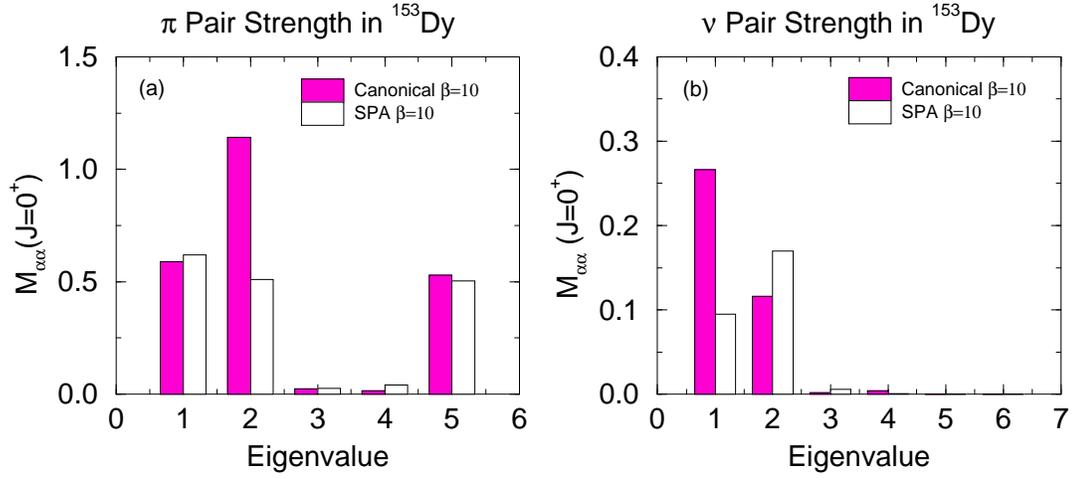,height=12cm}
\caption{(a) Pairing matrix eigenvalues for protons in $J^\pi=0^+$ 
for $^{153}$Dy at $T=.1$ MeV. (b) Pairing matrix eigenvalues for
neutrons in $J^\pi=0^+$ for $^{153}$Dy at $T=.1$ MeV.}
\label{fig:153j0}
\end{center}
\end{figure}

The pairing strengths in both proton and neutron channels has also been
computed (Fig.~\ref{fig:153j0}(a)-(b)).  The sum
of these eigenvalues, with no background subtraction, in p(n) $J=0^+$ channels
is $2.30(0.39)$ for the exact canonical solution and $1.70(0.27)$ in
SPA.  These values are stronger in the full canonical than in the SPA,
as would be expected from looking at Fig.~\ref{fig:153static}(b).  For the
protons, the difference in the eigenvalue sum is mostly due to
eigenvalue number $2$, where the full canonical eigenvalue is more
than twice the SPA result.  These eigenvalues are otherwise distributed 
very similarly in the full and SPA results.  A similar situation holds
for the neutrons, where the first eigenvalue for the full canonical
solution is more than double the SPA value.  SPA simply does not
produce the nuclear pair condensate revealed in the exact calculation.

\subsubsection{Cranking}
With cranking at $\beta=10$ MeV$^{-1}$, the sign for $^{153}$Dy is $0.69$ for
$\omega=0.05$ MeV, $0.52$ for $\omega=0.1$ MeV, and just $0.10$ for
$\omega=0.2$ MeV.  Recall that the sign for $\beta=10$ MeV$^{-1}$ in uncranked
$^{153}$Dy is $0.82$.  The moment of inertia, ${\cal I}_2$, is
$49.6\pm3$ $\hbar^2/$MeV in the limit $\omega\rightarrow 0$ for the
canonical calculation. 

\subsection{Level density}
\begin{figure}
\begin{center}
\hskip 0.25in
\psfig{figure=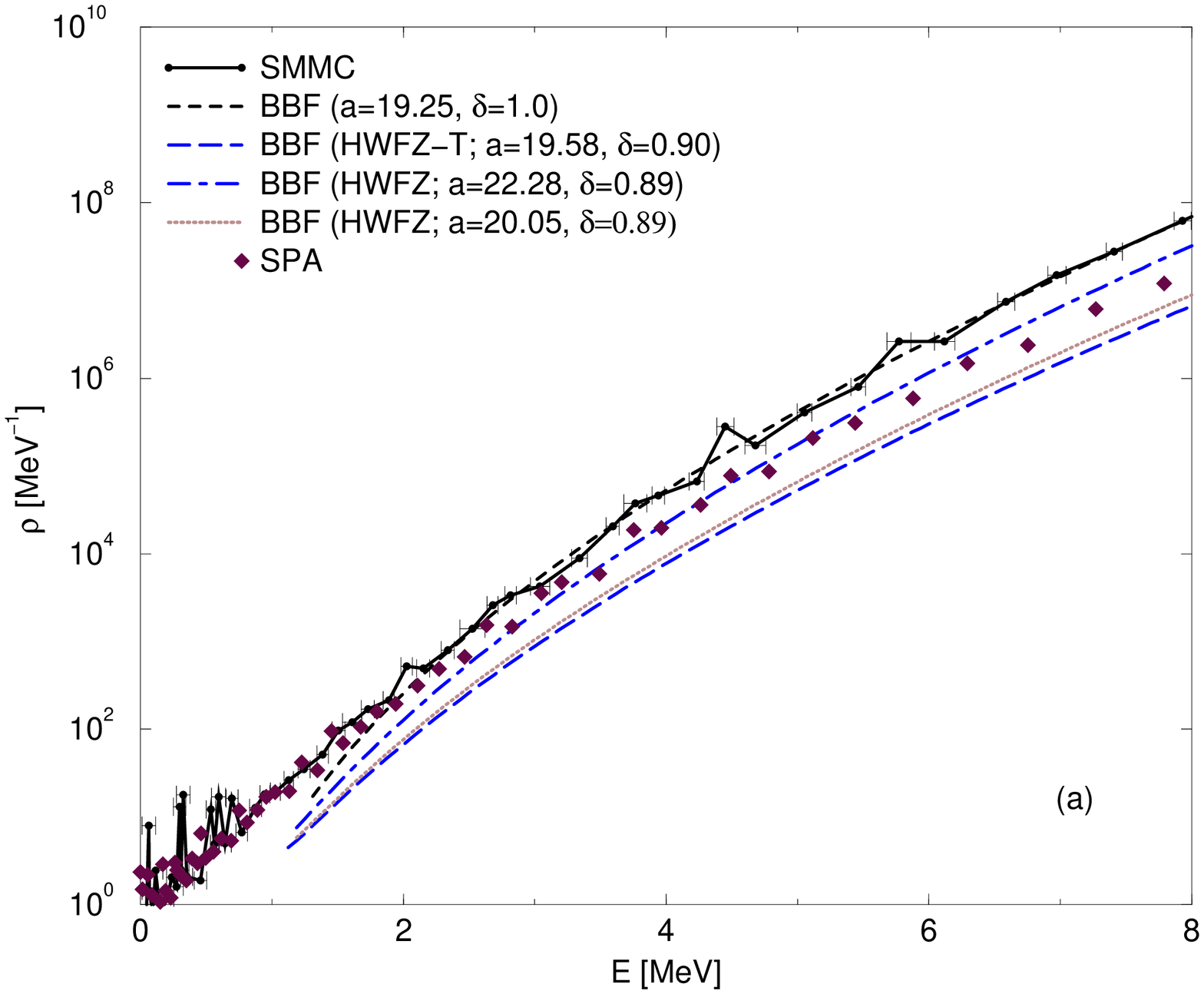,height=8cm,angle=-90}

\hskip 0.25in
\psfig{figure=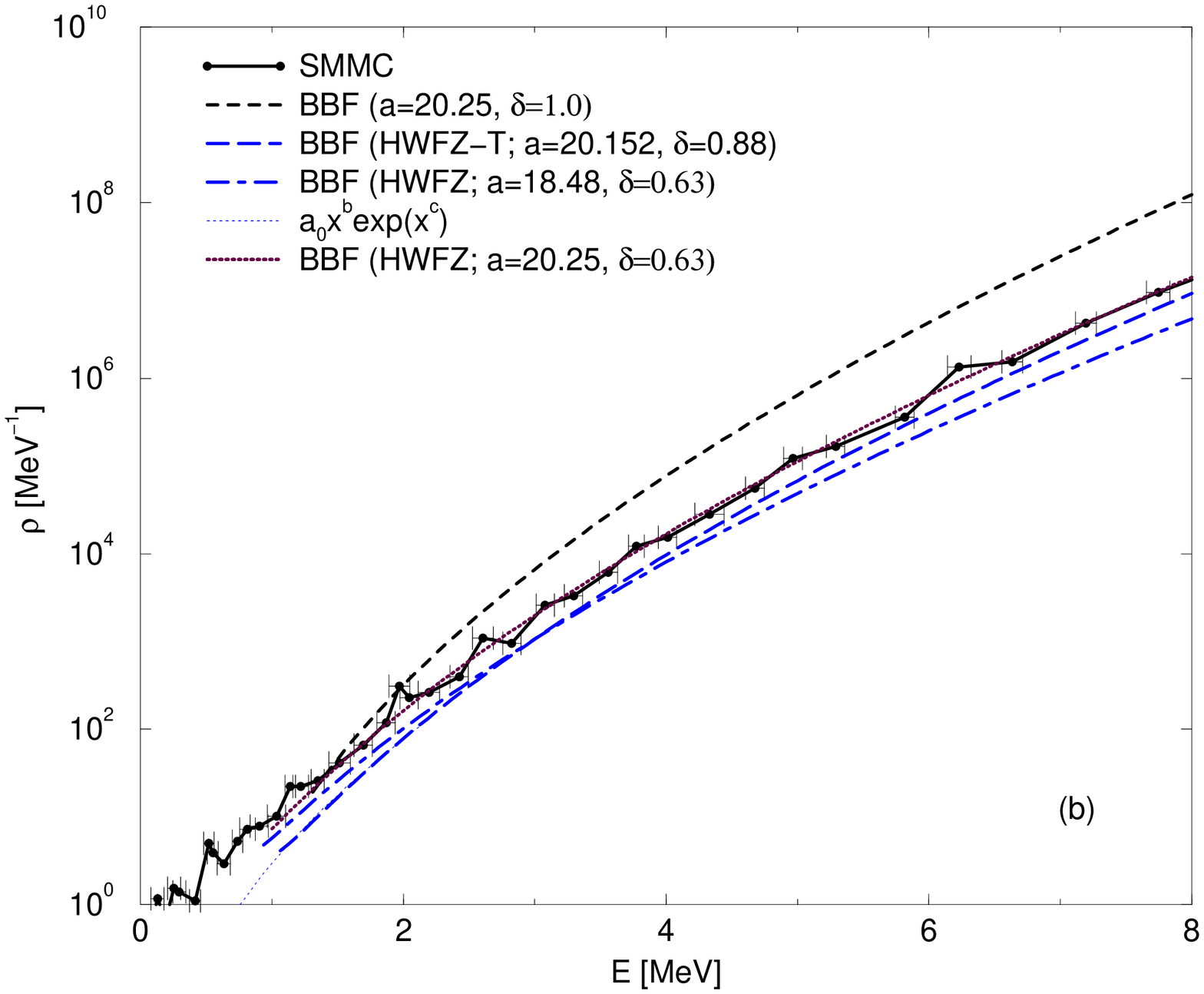,height=8cm,angle=-90}

\caption{(a) Level density for
$^{154}$Dy in SMMC shown with backshifted Fermi gas approximations.
BBF=standard backshifted Bethe formula.
HFWZ=parameterization of Holmes, Woosley, Fowler, and
Zimmerman, and HWFZ-T=HWFZ formula with Thielemann
parameters.  See text.  HWFZ with $a=22.28$ is for
spherical $^{154}$Dy parameters while $a=20.05$ is for deformed
$^{154}$Dy parameters.  SPA denotes the SPA result for the level
density from SMMC. (b) SMMC result for $^{162}$Dy compared with
the same backshifted Bethe formula approximations.}
\label{fig:dens}
\end{center}
\end{figure}

\begin{figure}
\begin{center}
\hskip 0.5in \psfig{figure=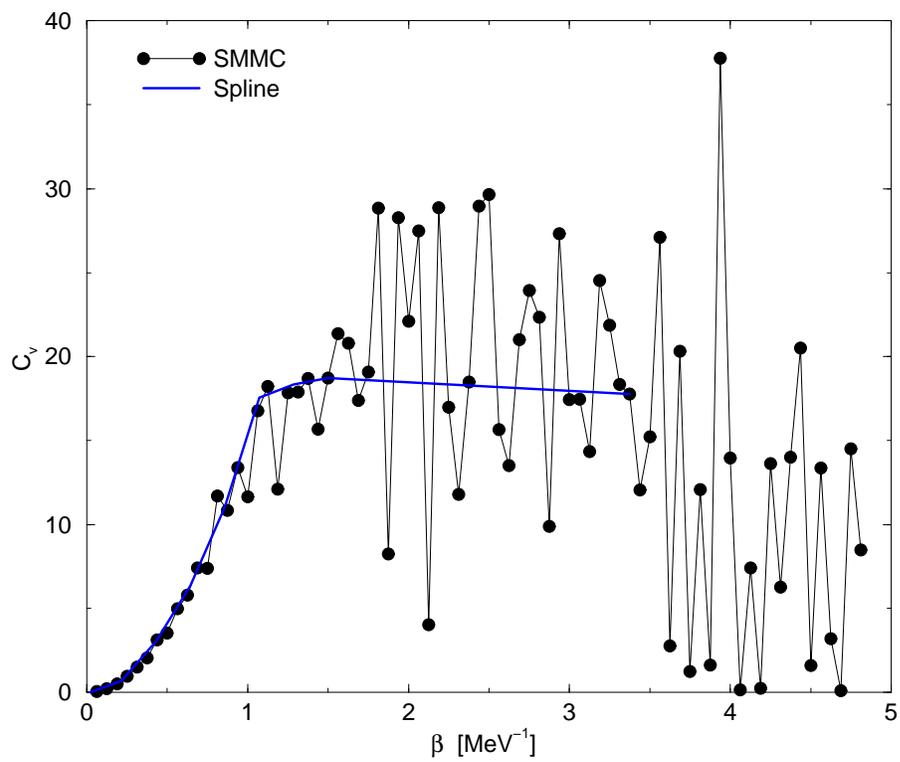,height=12cm,angle=-90}
\caption{Heat capacity in $^{154}$Dy.}
\label{fig:154cv}
\end{center}
\end{figure}

The level density results for $^{154}$Dy are shown in
Figs.~\ref{fig:dens}-~\ref{fig:154cv}.  $E(\beta)$ points are
calculated at intervals of $\Delta\beta=0.0625$ to execute the 
saddle point inversion to the level density $\rho(E)$
(Eq.~\ref{eq:rho}).  The level density (Fig.~\ref{fig:dens}(a)) is
compared with a few parameterizations of backshifted Fermi gas
formulas.  The $^{154}$Dy density is not directly compared with
experimental data since no measurements are available.  SMMC results
are not as accurate for low temperatures or small excitation energies
($E < 1$ MeV) since numerical errors tend to be larger there.  
This is not a serious concern since the saddle point
approximation itself is not really valid at the lowest energies
anyway.  For the lowest energies, density of states is best determined by
simple state counting from known experimental levels.

Three versions of Fermi gas density formulas are used.  The first,
labelled BBF with $a=19.25$ and $\delta=1.0$ in Fig.~\ref{fig:dens}(a), is the
classic Bethe 
formula~\cite{bethe}: 
\begin{equation}
\rho(E)=\frac{1}{12 a^{1/4}
(E-\delta)^{5/4}}\exp\left(2\sqrt{a(E-\delta)}\right) 
\end{equation}
The calculation for $^{154}$Dy was done with $a=A/8=19.25$ MeV$^{-1}$
and the energy is backshifted as $E-\delta$ for $\delta=1$ MeV for
an even-even nucleus.  This formula happens to agree quite well with the
SMMC prediction for the $^{154}$Dy density for energies above 2 MeV.
Notice that solutions to this formula 
will diverge as $E\rightarrow\delta$ for positive $\delta$, so the
result is only shown down to an energy where the density formula
yields a sensible result. 

Holmes, Woosley, Fowler, and Zimmerman (HWFZ) calculate
backshifted level densities as~\cite{a}:
\begin{equation}
\rho(E)=\frac{0.482}{A^{5/6}}(E-\delta)^{-3/2}\exp\left(2\sqrt{a(E-\delta)}\right)
\end{equation}
The $a$ parameters for HWFZ can depend on whether the nucleus is
deformed or not.  For $^{154}$Dy, $\delta=0.89$ MeV and $a=22.28$
MeV$^{-1}$ for spherical parameters 
and $a=20.05$ for the deformed parameterization.  The spherical HWFZ
curve is always slightly low and the magnitude tails off to quickly
below $E=2.5$ MeV as compared with the SMMC result.  
HWFZ (spherical) is too small by a factor 2.5 at $E=10$ MeV and too low by
a factor of 4 at $E=1.5$ MeV.  HWFZ (deformed), which has a smaller
$a$ parameter, is clearly a worse fit.  It is worse than an order of
magnitude smaller at $E=10$ MeV and is six times smaller at $E=1.5$
MeV.  The typical BBF $a$ parameter, $A/8$, is $19.25$ MeV$^{-1}$ for $A=154$.
This is smaller than the HWFZ(deformed) density parameter and would
make the fit even worse.

Thielemann has modeled the parameters $\delta$ and $a$ slightly
differently than HWFZ.  In this paper, these are called HWFZ-T
parameters.  He has taken 
$\delta$ as 
\begin{equation}
\delta=\Delta(Z,N)-10/A
\end{equation}
with
\begin{eqnarray}
\Delta(Z,N)&=&12/\sqrt{A} \qquad even-even \\
           &=&-12/\sqrt{A}\qquad odd-odd \\
           &=& 0 \qquad\qquad odd
\end{eqnarray}
He obtained the density parameter $a$ from a fit to experimental
densities at one neutron separation energy~\cite{rohr,Truran91}.  For
$^{154}$Dy, this gives $\delta=0.90$ and $a=19.58$.  The HWFZ-T level
density is somewhat lower than the calculated SMMC density in
$^{154}$Dy at all energies, but the slope agrees pretty well with the
SMMC calculation.  The HWFZ-T magnitude is lower by a factor 15 at
$E=10$ MeV for $^{154}$Dy and a factor 6 for $E\approx1.5$ MeV.

Certainly for the case of $^{154}$Dy, the most naive Bohr-Mottelson
Fermi gas formula works much better than the more carefully developed
parameterizations of HWFZ and Thielemann.  This
serves as an example of the utility of more
realistic SMMC calculations to determine nuclear level densities.

From the specific heat (Fig.~\ref{fig:154cv}) and the known
$E $ vs. $\beta$, the $^{154}$Dy density
calculation is expected to be valid up to $10-15$ MeV excitation
before finite model space effects set in.  The specific heat will increase with
increasing temperature.  Eventually, however, the model space will
become exhausted as the valence particles
are all promoted as high in energy as possible within the finite
space.  The turnover point where $C_v$ stops decreasing is taken as
the limit of validity for the calculation.  An inert core is assumed
here at all times.

The SPA level density for $^{154}$Dy has also been calculated and
compared with SMMC (Fig.~\ref{fig:dens}(a)).  
The SPA level density agrees
well with SMMC for low excitation energies, but is consistently lower
for energies above 4 MeV.  Recall the SPA energy $E$ vs. $\beta$
(Fig.~\ref{fig:static}(d)) agrees with the full SMMC at high
temperatures, but never cools completely to the SMMC value for lower
temperatures.  Thus $\Delta E$ from $T=0$ to $T=\infty$ is smaller in
SMMC and this difference of a couple MeV makes a perceptible
difference in the level density.  At $E=10$ MeV, the SPA density is
smaller by a factor of 4.

\begin{figure}
\begin{center}
\hskip 0.5in \psfig{figure=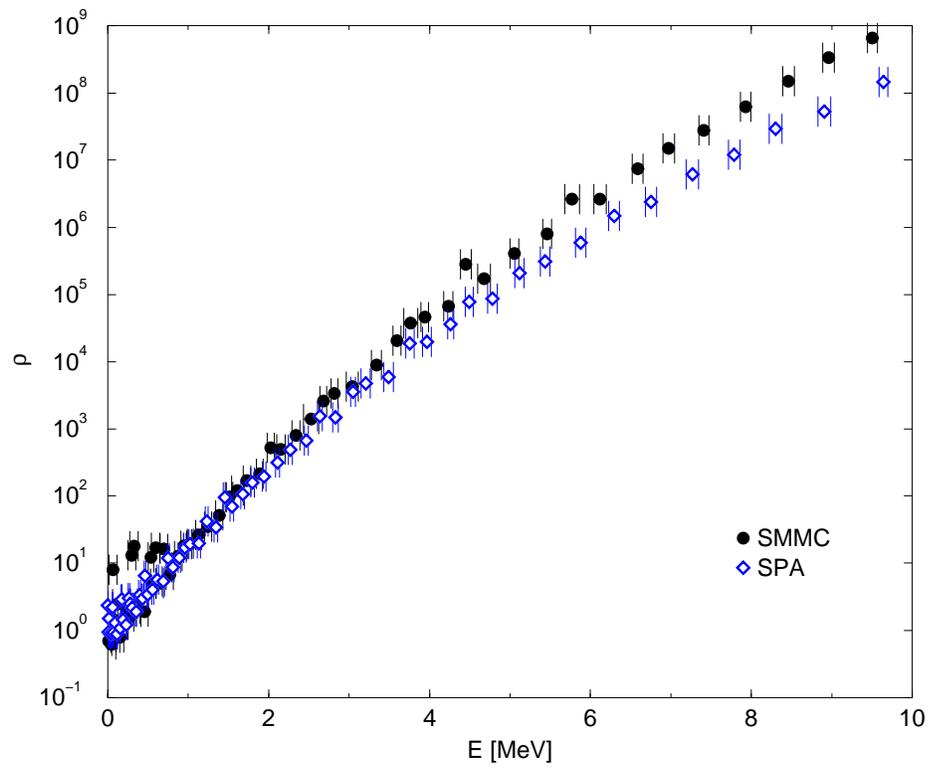,height=12cm,angle=-90}
\caption{SPA heat capacity in $^{154}$Dy.}
\label{fig:154spacv}
\end{center}
\end{figure}

The heat capacity can be found in Fig.~\ref{fig:154spacv}.  This looks
similar to the full SMMC calculation except the magnitude of $C_v$ is
smaller except for the lowest temperatures (highest $\beta$).  The
heat capacity has a sharp dropoff below $E=1$ MeV for both SPA and
full SMMC solutions.  The heat curve implies that SPA should be valid
for up to about $13$ MeV excitation.  However, the SPA density clearly diverges
from SMMC well before this limit.

Similar calculations are shown for $^{162}$Dy in
Fig.~\ref{fig:dens}(b).  For the more deformed
$^{162}$Dy, the HWFZ-T formula works comparatively well as Fermi gas
estimates go, but is still off by a factor 3 near $E=1$ MeV and factor
1.4 near $E=10$ MeV.  HWFZ in $^{162}$Dy is better than HWFZ-T at low
energies, but is clearly worse at higher energies.  It is within a
factor 2 of SMMC for $E=1$ MeV and smaller than SMMC by a factor of 3
at $E=10$ MeV.  In contrast to $^{154}$Dy, HWFZ fits very well for $^{162}$Dy
using $a=A/8=20.25$ MeV$^{-1}$.  The simple backshifted Bethe formula
fails badly here, however, especially for higher energies. 

We determined that the level density calculation for this isotope is
valid up to excitations of $15-20$ MeV.  This is slightly higher than
the valid range for the density in $^{154}$Dy. 

\begin{figure}
\begin{center}
\hskip 0.5in \psfig{figure=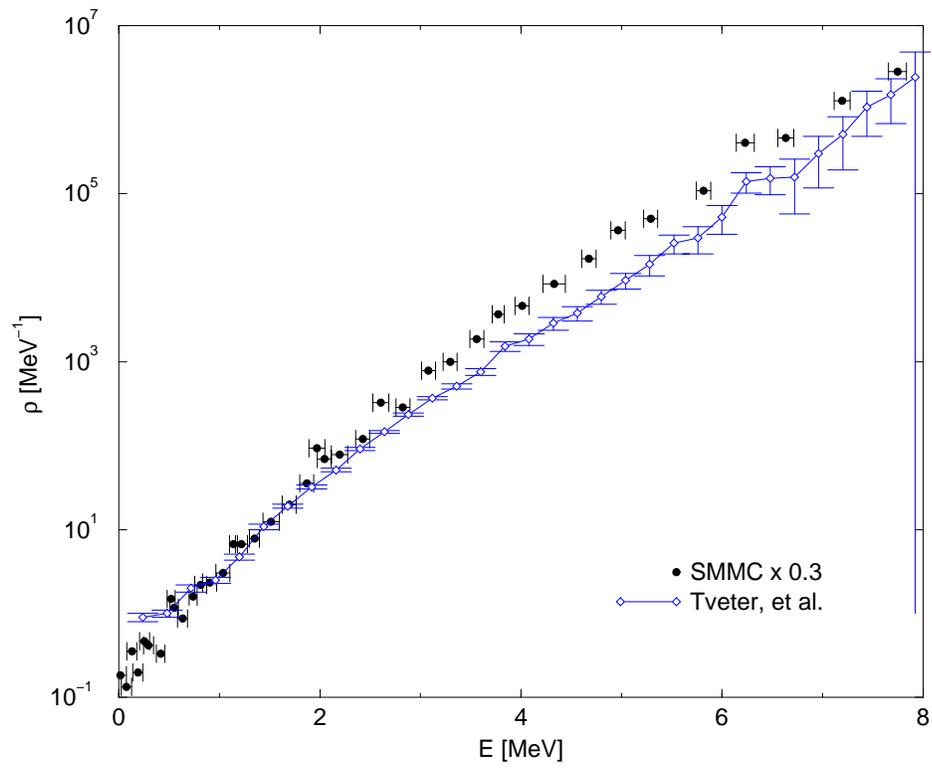,height=12cm,angle=-90}
\caption[SMMC density vs. data in $^{162}$Dy]{SMMC density vs. data in
$^{162}$Dy.}
\label{fig:162rhovsdat}
\end{center}
\end{figure}

The comparison of SMMC density in $^{162}$Dy with the Tveter, et
al~\cite{Tveter96} data is displayed in Fig.~\ref{fig:162rhovsdat}.
The experimental method of Tveter, et al. can reveal fine structure,
but does not determine the absolute 
density magnitude.  The SMMC calculation is scaled to
facilitate comparison.  In this case, the scale factor has been chosen to
make the curves agree at lower excitation energies.  From $1-3$ MeV,
the agreement is very good.  From $3-5$ MeV, the SMMC density
increases more rapidly than the 
data.  This deviation from the data cannot be accounted for by
statistical errors in either the calculation or measurement.  Near 6
MeV, the measured density briefly flattens before increasing and this
also appears in the calculation, but the measurement errors are larger
at that point.  

The measured density includes all states
included in the theoretical calculation plus some others, so that one
would expect the measured density to be greater than or equal to the
calculated density and never smaller.  We may have instead chosen our
constant to match the densities for moderate excitations and let the
measured density be higher than the SMMC density for lower energies
(1-3 MeV).  

Comparing structure between SMMC and data is difficult
for the lowest energies due to statistical errors in the calculation
and comparison at the upper range of the SMMC calculation, i.e.,
$E\approx15$ MeV, is unfortunately impossible since the data only
extend to about $8$ MeV excitation energy.  

\begin{figure}
\begin{center}
\hskip 0.5in \psfig{figure=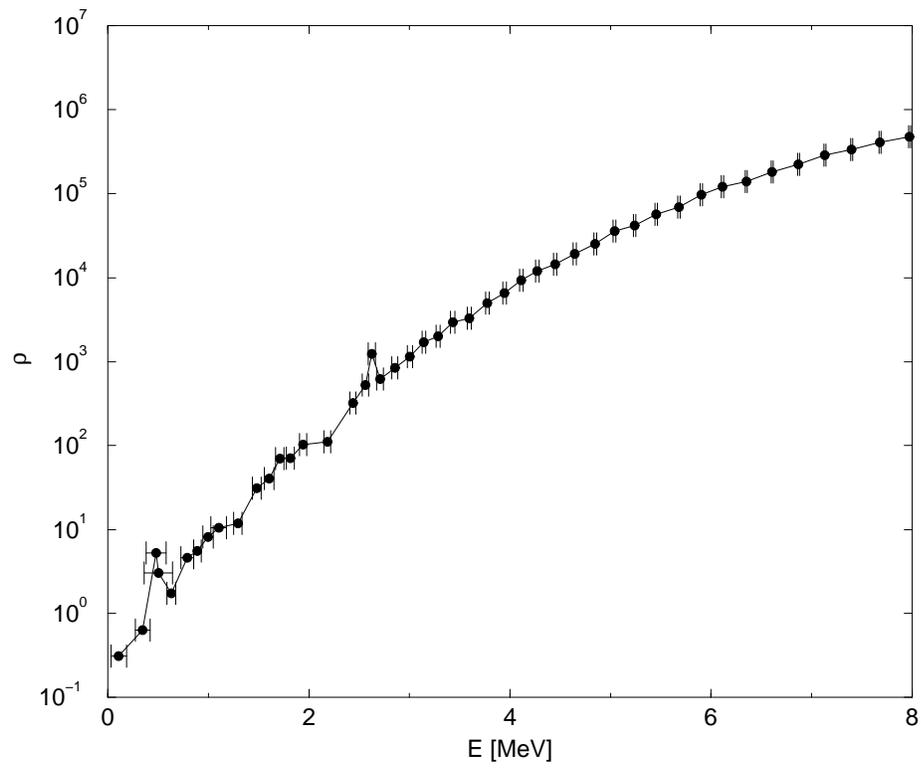,height=12cm,angle=-90}
\caption[SMMC density in $^{140}$Ba]{SMMC density in
$^{140}$Ba.}
\label{fig:ba140rho}
\end{center}
\end{figure}

\begin{figure}
\begin{center}
\hskip 0.5in \psfig{figure=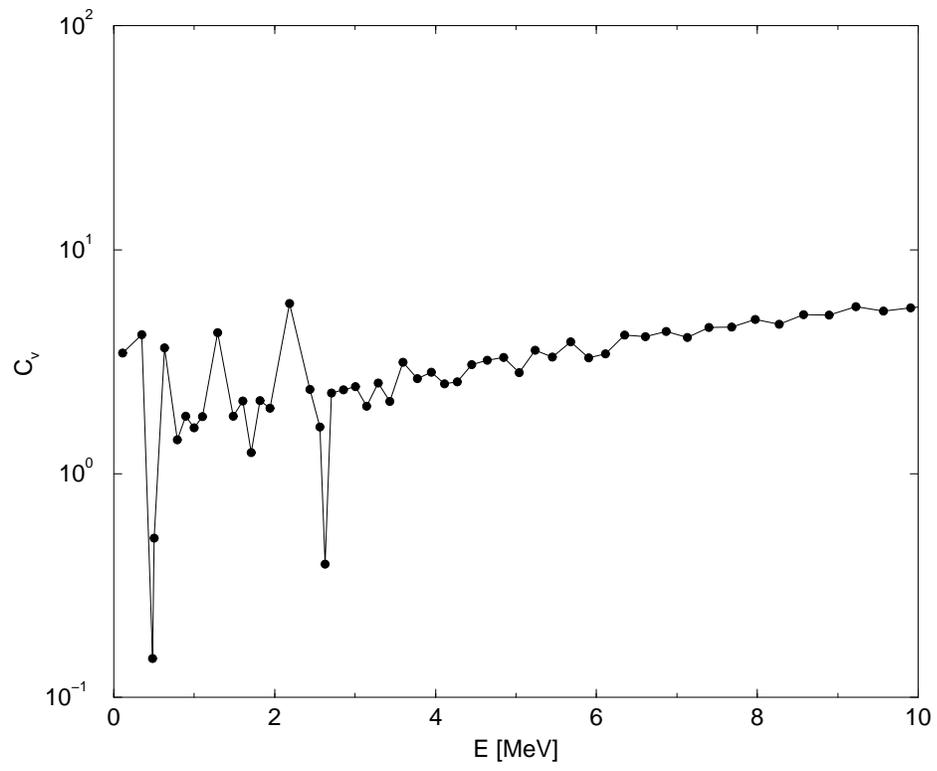,height=12cm,angle=-90}
\caption[SMMC heat capacity in $^{140}$Ba]{SMMC heat capacity in
$^{140}$Ba.}
\label{fig:ba140cv}
\end{center}
\end{figure}

Level density information has also been calculated for the lighter
nearer closed shell nucleus $^{140}$Ba.  This was done to investigate
possible systematic differences in level densities.  Its level density
is shown in Fig.~\ref{fig:ba140rho} and the specific heat in
Fig.~\ref{fig:ba140cv}.  Unlike the dysprosiums, the calculated heat
capacity curve in $^{140}$Ba is very flat.

\section{Summary}
The work has systematically laid the groundwork for applying
the shell model in rare earths.  Previous applications have been
plagued by severely truncated model spaces.  An advantage of being
able to explore exact shell model solutions in more expansive 
model spaces is to explain in fundamental ways behaviors such as band
crossings and pair correlations which have been previously understood
from phenomenological models. 

The static path approximation for this phenomenological pairing plus 
quadrupole model works well for calculating deformation and relative
energy differences between ground states of different isotopes
regardless of deformation.  Additionally, deformations are well
determined in SPA for quadrupole coupling strengths even a factor of three
larger or smaller than the Kumar-Baranger mean field values.

The SPA results for the pairing energy and pair gaps are not as good,
however, and the discrepancy is worse for increasing pair strengths.
SPA also overestimates the low spin moments of inertia.  However, 
SPA does produce the $\nu i_{13/2}$ band crossing at the predicted
spin for $^{156}$Dy.  SPA does not produce the ground state nuclear
pair condensate and pair gap, hence the discrepancies in energy and
moments of inertia.

Deformations in the canonical ensemble with Kumar-Baranger parameters
agree with both Kumar-Baranger and experimental results for isotopes
tested that appear in their paper~\cite{Kumar-Baranger}, but
deformations calculated for some other nuclei do not.  Also, the
deformation wells in Dysprosiums with $A\geq156$ are very deep at low
temperatures, i.e., below $T\sim 0.25$ MeV.  There is some question whether
the shape plots at such low temperatures are reliable.  However,
$^{152}$Dy is clearly too deformed in its ground state and the 
calculated $B(E2)$ strengths require a reduced fitted
effective charge for $A\leq 154$, indicating that the
light Dysprosiums are excessively deformed in their ground states for
this Hamiltonian. 

\acknowledgements
J. White is very grateful to T. L. Khoo, K. Langanke, W. Nazarewicz, and
P. Vogel for useful discussions.  This work was supported in part by
the National Science Foundation,   
Grants No. PHY-9722428, PHY-9420470, and PHY-9412818.  This work was
also supported in part through grant DE-FG02-96ER40963 
from the U.S. Department of Energy.  Oak Ridge National Laboratory
(ORNL) is managed by Lockheed Martin Energy Research Corp. for the
U.S. Department of Energy under contract number 
DE-AC05-96OR22464.  We also acknowledge use of the CACR parallel
computer system operated by Caltech and use of MHPCC SP2 systems.


\newpage
\newpage

\newpage

\typeout{Ba: Qp^2=67.4, Qn^2=25.1, Qv^2=51.4; Q^2=133.4}
\typeout{4.18 MeV; 10 contours}

\newpage

\newpage
\typeout{check Thielemann article in truran91}

\begin{table}
\begin{center}
\caption[SMMC $B(E2)$ vs. measured $B(E2)$]{SMMC
$B(E2)$ vs. measured
$B(E2;2_1^+\rightarrow 0_1^+)$ with specified effective charges.
$B(E2)$ in W.U.  Errors are statistical Monte Carlo
sampling errors.  Column 6, $B(E2)$ for SMMC/KB $e_p(n)$, is the
$B(E2)$ obtained from SMMC calculated quadrupole moments with
Kumar-Baranger effective charges.}  
\vspace{0.1in}
\label{table:be2}
\begin{tabular}{|c|c||c|c|c|c|c|}\hline
A & N & $e_p$ & $e_n$ & $B(E2)$ & $B(E2)$ & $B(E2)$ \\
  &   &       &       & SMMC    & SMMC/KB $e_{p(n)}$ & Expt \\ \hline\hline
152 & 86 & 1 & 0 & $13\pm0.2$ & N/A & 13 \\
154 & 88 & 1.5 & 0.5 & $97\pm0.2$ & N/A & 97 \\
156 & 90 & 1.75 & 0.75 & $146\pm0.6$ & $126.4\pm0.5$ & 146 \\
162 & 96 & 1.75 & 0.75 & $199\pm0.7$ & $150.1\pm0.5$ & 199 \\ \hline
\end{tabular}
\end{center}
\end{table}

\begin{table}
\begin{center}
\caption[Spherical $E(2_1^+)$ values from Kumar-Baranger]{Some spherical
  $E(2_1^+)$ values from Kumar-Baranger~\cite{Kumar-Baranger}.
  Energies in keV.}  
\vspace{0.1in}
\label{table:kbspherbe2}
\begin{tabular}{|c|||c|c|}\hline
  & $E(2_1^+)_{exp}$ & $E(2_1^+)_{th}$ \\ \hline\hline
$^{138}$Ba & 1438 & 2767 \\
$^{140}$Ba & 602  & 2006 \\
$^{140}$Ce & 1596 & 2531 \\
$^{142}$Ce & 641 & 1772 \\
$^{144}$Ce & 397 & 1095 \\ \hline
\end{tabular}
\end{center}
\end{table}

\end{document}